\documentclass[aps,prd,onecolumn,superscriptaddress,nofootinbib,longbibliography,showkeys,floatfix]{revtex4-2}

\usepackage{amsmath,amssymb,bm,graphicx,booktabs,array}
\usepackage{mathrsfs}
\usepackage{hyperref}
\hypersetup{
  colorlinks=true,
  citecolor=blue,
  linkcolor=blue,
  urlcolor=blue,
  pdftitle={Coupled Gravitoelectromagnetic Response of a Magnetically Supported Generalized Hayward Black Hole in Nonlinear Electrodynamics},
  pdfauthor={Anirudh Pradhan, K. Ghaderi, M. Zeyauddin},
  pdfsubject={Coupled black hole perturbations and wave conversion in nonlinear electrodynamics},
  pdfkeywords={nonlinear electrodynamics, coupled perturbations, quasinormal modes, wave conversion, greybody factors}
}

\newcommand{\Lag}{\mathcal{L}}
\newcommand{\F}{\mathcal{F}}
\newcommand{\eps}{\varepsilon}
\newcommand{\kap}{\kappa}
\newcommand{\vk}{\varkappa}
\newcommand{\bV}{\bm{V}}
\newcommand{\bPsi}{\bm{\Psi}}
\newcommand{\bR}{\bm{\mathcal R}}
\newcommand{\bT}{\bm{\mathcal T}}
\newcommand{\bA}{\bm{\mathcal A}}
\newcommand{\id}{\bm{I}}

\begin{document}
\raggedbottom

\title{Coupled Gravitoelectromagnetic Response of a Magnetically Supported Generalized Hayward Black Hole in Nonlinear Electrodynamics}

\author{Anirudh Pradhan}
\affiliation{Centre for Cosmology, Astrophysics and Space Science (CCASS), GLA University, Mathura 281406, U.P., India}

\author{K.~Ghaderi}
\email{k.ghaderi@iau.ac.ir}
\affiliation{Department of Physics, Mari.C., Islamic Azad University, Marivan, Iran}

\author{M.~Zeyauddin}
\affiliation{Department of General Studies (Mathematics), Jubail Industrial College, Jubail 31961, Saudi Arabia}

\author{A.~Gulhane}
\affiliation{PTC Software Inc. India}

\begin{abstract}
We construct the exterior coupled gravitational and electromagnetic response of a magnetically supported generalized Hayward black hole within a specified nonlinear electrodynamic completion. Both parity sectors reduce to canonical two channel wave systems with explicit potentials and the correct Schwarzschild and effective Reissner--Nordstr\"om limits. We compute the quasinormal spectrum, metric and optical characteristics, extremal throat weights, quadrupolar wave conversion, finite bandwidth transfer, and coherent absorption eigenchannels while verifying exterior hyperbolicity and flux conservation. An effective charged operator captures the leading weak core geometry and mode mixing; after its subtraction, the residual hierarchy is consistent with cubic spectral corrections in the primarily electromagnetic branch and quartic corrections in the primarily gravitational branch and principal scattering observables. The fundamental quadrupolar branches become intrinsically linewidth separated near $\chi\simeq0.70$, whereas parity splitting remains subleading. Metric and optical characteristic families separate monotonically toward extremality. At $\chi=0.9$, positive parity reflected conversion reaches $45.8\%$ and remains $40.8\%$ after moderate bandwidth averaging, while coherent incident combinations yield $98.4\%$ bright and $1.3\%$ dark absorption. These independently evaluated observables provide mutually consistent diagnostics of the same nonlinear electrodynamic operator.
\end{abstract}


\maketitle

\section{Introduction}
\label{sec:intro}

Black hole perturbation theory maps a gravitational model to complementary dynamical observables. Complex frequency poles determine the intrinsic ringdown spectrum, characteristic surfaces control the short wavelength limit, and the real frequency response fixes reflection, transmission, absorption, and channel conversion. These quantities probe one linear operator and should therefore be calculated with a common matter completion and normalization \cite{Chandrasekhar1983,BertiCardosoStarinets2009,KonoplyaZhidenko2011,BertiEtAl2026}.

Regular black hole geometries replace the Schwarzschild curvature singularity by a finite density core while retaining an asymptotically flat exterior \cite{AyonBeatoGarcia1998,Hayward2006,Bronnikov2001}. Generalized Hayward constructions encompass several controlled core profiles and have been studied at the geometric and test field levels \cite{DuttaRoyKar2022,KudryavcevLingVertogradov2026}. A prescribed lapse, however, is not a complete perturbative model: the supporting matter Lagrangian enters the linearized equations through its constitutive derivatives. Test field calculations therefore do not determine the coupled gravitational and electromagnetic response of the self gravitating system.

Nonlinear electrodynamics supplies a magnetic completion for several regular black hole geometries \cite{AyonBeatoGarcia2000,Bronnikov2001,MorenoSarbach2003}. The perturbation problem differs from the Einstein Maxwell case because gravitational and electromagnetic fluctuations are coupled in both parity sectors and the electromagnetic principal part is governed by an optical geometry \cite{NovelloEtAl2000,SchellstedeEtAl2016,ToshmatovEtAl2018,ToshmatovEtAl2018Polar,TomizawaSuzuki2023}. Gauge invariant formulations and exterior stability criteria are available for self gravitating nonlinear electrodynamic black holes \cite{MorenoSarbach2003,DaghighGreen2022,Breton2005,NomuraEtAl2020}, while constitutive nonlinearities are known to break the parity isospectrality of Reissner--Nordstr\"om black holes \cite{ChaverraEtAl2016,NomuraYoshida2022}. Coupled gravitoelectromagnetic quasinormal modes have been computed for magnetic regular black holes \cite{MengZhang2023}; recent Hayward studies have also addressed axial gravitational modes and greybody observables \cite{Malik2025,WuCaiXie2025,BolokhovSkvortsova2026}, and related nonlinear electrodynamic geometries have been analyzed spectrally \cite{LiangEtAl2026}. The present work differs by carrying a specified magnetic completion and its associated canonical two channel operator through the complex frequency spectrum, characteristic limits, real frequency conversion, finite bandwidth transfer, and coherent absorption.

The real frequency response contains information not encoded in quasinormal frequencies alone. Charged black holes convert incident gravitational radiation into electromagnetic radiation and conversely \cite{OuldElHadjDolan2022}. In a two channel problem, the canonical scattering matrix is the natural object: its off diagonal elements determine conversion, while the eigenvalues of the absorption matrix identify coherent combinations that couple strongly or weakly to the horizon.

We perform a unified exterior analysis of the specific generalized Hayward member defined in Eq.~\eqref{eq:mass} and supported by the magnetic nonlinear electrodynamic completion reconstructed below. The calculation includes both parity sectors, continuous branch tracking, effective Reissner--Nordstr\"om subtraction, extremal throat reduction, quadrupolar scattering and packet transfer, and coherent greybody eigenchannels. The resulting chain is
\begin{equation}
\begin{aligned}
 \text{background completion}&\rightarrow\text{coupled operator}\\
 &\rightarrow\text{spectrum and characteristics}\\
 &\rightarrow\text{real frequency response}.
\end{aligned}
\end{equation}

All stability statements are restricted to the domain of outer communication. We verify exterior kinetic and hyperbolicity conditions, regularity of the canonical system, positive sampled potential eigenvalues, flux conservation, and decaying modes in the investigated spectral region. These results are logically distinct from questions concerning the central region or the placement of a regular solution within a broader singular family \cite{DeFeliceTsujikawa2025,HuangRao2025,BokulicJuricSmolic2026}.

The paper is organized as follows. Section~\ref{sec:background} fixes the background and reconstructs its magnetic nonlinear electrodynamic completion. Section~\ref{sec:operator} defines the canonical perturbation systems and exterior consistency tests. Section~\ref{sec:rn} develops the effective charged expansion. Sections~\ref{sec:qnm}--\ref{sec:throat} present the quasinormal, characteristic, and extremal throat results. Section~\ref{sec:scattering} gives the quadrupolar scattering matrix, conversion, packet transfer, and coherent absorption eigenchannels. Section~\ref{sec:closure} summarizes cross observable checks. 
Section~\ref{sec:discussion} gives the physical interpretation, and Sect.~\ref{sec:conclusion} summarizes the conclusions. 

\section{Background and specified nonlinear electrodynamic completion}
\label{sec:background}

\subsection{Geometry and horizon branch}

We use geometrized units \(G=c=\hbar=k_{\rm B}=1\) and signature \((-+++)\).  The static line element is
\begin{equation}
 ds^{2}=-f(r)dt^{2}+\frac{dr^{2}}{f(r)}+r^{2}d\Omega^{2},
 \qquad
 f(r)=1-\frac{2m(r)}{r},
 \label{eq:metric}
\end{equation}
with mass function
\begin{equation}
 m(r)=M\left[1-\frac{\beta}{(r^{3}+\beta^{3})^{1/3}}\right].
 \label{eq:mass}
\end{equation}
The parameter \(M\) is the ADM mass and \(\beta>0\) is the core scale. Equation~\eqref{eq:mass} specifies one magnetically supported member of the generalized Hayward class rather than the full multiparameter family \cite{DuttaRoyKar2022,KudryavcevLingVertogradov2026}.  We define
\begin{equation}
 \eps=\frac{\beta}{M},
 \qquad
 x=\frac{r}{\beta}.
 \label{eq:dimensionless}
\end{equation}
The horizon equation can be written as
\begin{equation}
 \eps=2\,\frac{1-(1+x^{3})^{-1/3}}{x}.
 \label{eq:horizon}
\end{equation}
The maximum of the right hand side determines the extremal configuration,
\begin{align}
 x_{\rm e}&=1.734947084082,\qquad
 \eps_{\rm e}=0.526022945016,\\
 \frac{r_{\rm e}}{M}&=0.912621974616.
 \label{eq:extremal}
\end{align}
We parameterize the black hole branch by
\begin{equation}
 \chi=\frac{\eps}{\eps_{\rm e}},
 \qquad 0<\chi\leq1.
 \label{eq:chi}
\end{equation}
The outer surface gravity and Hawking temperature are
\begin{equation}
 \kap_{+}=\frac12 f'(r_{+}),\qquad T_{\rm H}=\frac{\kap_{+}}{2\pi}.
 \label{eq:surfacegravity}
\end{equation}

At large radius,
\begin{equation}
 f(r)=1-\frac{2M}{r}+\frac{2M\beta}{r^{2}}
 -\frac{2M\beta^{4}}{3r^{5}}+O(r^{-8}).
 \label{eq:asymptoticf}
\end{equation}
The \(r^{-2}\) term anticipates the effective charged correspondence developed in Sec.~\ref{sec:rn}.

\subsection{Action and magnetic reconstruction}

We use the nonlinear electrodynamic normalization standard in the coupled magnetic perturbation formalism \cite{MorenoSarbach2003,ChaverraEtAl2016}. The action and electromagnetic invariant are
\begin{equation}
 S=\frac{1}{16\pi}\int d^{4}x\sqrt{-g}
 \left[R-4\Lag(\F)\right],
 \qquad
 \F=\frac14 F_{\mu\nu}F^{\mu\nu}.
 \label{eq:action}
\end{equation}
The field equations are
\begin{align}
 G_{\mu\nu}&=
 2\Lag_{\F}F_{\mu\alpha}F_{\nu}{}^{\alpha}
 -2g_{\mu\nu}\Lag,
 \label{eq:einstein}\\
 \nabla_{\mu}\left(\Lag_{\F}F^{\mu\nu}\right)&=0.
 \label{eq:nedfield}
\end{align}
For the magnetic monopole
\begin{equation}
 F_{\theta\phi}=Q_{\rm m}\sin\theta,
 \qquad
 \F=\frac{Q_{\rm m}^{2}}{2r^{4}}>0,
 \label{eq:magnetic}
\end{equation}
the \(tt\) equation gives \(m'=r^{2}\Lag\).  Matching the Maxwell weak field limit fixes
\begin{equation}
 Q_{\rm m}^{2}=2M\beta.
 \label{eq:qmbeta}
\end{equation}
Defining
\begin{equation}
 y=\left(\frac{\beta}{r}\right)^{3}
 =\beta^{3}\left(\frac{2\F}{Q_{\rm m}^{2}}\right)^{3/4},
 \label{eq:y}
\end{equation}
the on-shell single-invariant completion along the monotonic magnetic branch is
\begin{equation}
 \Lag(\F)=\F(1+y)^{-4/3}.
 \label{eq:L}
\end{equation}
Its constitutive derivatives are
\begin{align}
 \Lag_{\F}&=(1+y)^{-7/3},
 \label{eq:LF}\\
 \Lag_{\F\F}&=-\frac{7y}{4\F}(1+y)^{-10/3},
 \label{eq:LFF}\\
 \Lag_{\F\F\F}&=
 \frac{7y(11y+1)}{16\F^{2}}(1+y)^{-13/3}.
 \label{eq:LFFF}
\end{align}
Equations~\eqref{eq:action}--\eqref{eq:LFFF} fix the constitutive functions sampled throughout the exterior and remove the ambiguity of perturbing the same lapse with an unrelated matter model. Across the solution family, the functional form in Eq.~\eqref{eq:L} is retained, while its scale parameters obey the background relation in Eq.~\eqref{eq:qmbeta}.

\subsection{Exterior principal conditions}

The electromagnetic characteristic factor for the magnetic background is
\begin{equation}
 \vk
 =1+\frac{2\F\Lag_{\F\F}}{\Lag_{\F}}
 =\frac{2-5y}{2(1+y)}.
 \label{eq:kappaopt}
\end{equation}
The exterior kinetic and hyperbolicity requirements are
\begin{equation}
 \Lag_{\F}>0,\qquad \vk>0.
 \label{eq:hyperbolicity}
\end{equation}
The first condition is automatic.  The zero of \(\vk\) occurs at \(y=2/5\), inside the event horizon throughout the black hole branch.  At extremality,
\begin{equation}
 \vk_{\rm e}=0.437503643250.
 \label{eq:kappae}
\end{equation}
The sufficient even parity condition inherited from the gauge invariant stability analysis is \(0<f\vk\leq3\) in the domain of outer communication \cite{MorenoSarbach2003,NomuraEtAl2020}. For the present branch, \(0<f<1\) and \(0<\vk\leq1\) outside the horizon, so this inequality follows directly once Eq.~\eqref{eq:hyperbolicity} is satisfied. The canonical denominators remain nonzero throughout the same domain. These statements concern \(r\geq r_{+}\) and are not extrapolated to the central region.

\section{Canonical coupled perturbation systems}
\label{sec:operator}

Perturbations are decomposed into spherical harmonics and parity sectors \(\mathcal P=-,+\).  After eliminating constraints and fixing the canonical normalization, each sector is represented by
\begin{equation}
 \bPsi_{\mathcal P}
 =
 \begin{pmatrix}
 \Psi_{\mathcal P}^{(g)}\\
 \Psi_{\mathcal P}^{(e)}
 \end{pmatrix},
 \label{eq:mastervector}
\end{equation}
where the asymptotic components reduce continuously to gravitational and electromagnetic channels.  For \(\ell\geq2\),
\begin{equation}
 \left[
 \partial_{t}^{2}-\partial_{r_{*}}^{2}
 \right]\bPsi_{\mathcal P}
 +
 \bV_{\mathcal P}(r)\bPsi_{\mathcal P}=0,
 \qquad
 \frac{dr_{*}}{dr}=\frac1f.
 \label{eq:master}
\end{equation}
The potential matrix is real and symmetric,
\begin{equation}
 \bV_{\mathcal P}
 =
 f\begin{pmatrix}
 U_{\mathcal P}^{gg} & U_{\mathcal P}^{ge}\\
 U_{\mathcal P}^{ge} & U_{\mathcal P}^{ee}
 \end{pmatrix},
 \label{eq:V}
\end{equation}
and is obtained by substituting Eqs.~\eqref{eq:mass} and \eqref{eq:L}--\eqref{eq:LFFF} into the gauge invariant Einstein nonlinear electrodynamic perturbation equations \cite{MorenoSarbach2003,DaghighGreen2022}. The explicit odd and even parity entries, including all field redefinitions and radial derivative terms used in the numerical calculations, are given in Appendix~\ref{app:operator}. Direct evaluation of these expressions yields the matrices and eigenvalues used throughout the analysis.

The canonical form is essential. It makes the conserved Wronskian and the flux metric Euclidean, ensures that the scattering probabilities are basis independent under constant orthogonal rotations, and permits a direct comparison between the complex frequency and real frequency problems. In the Maxwell limit this construction is continuously connected to the Moncrief formulation of Reissner--Nordstr\"om perturbations \cite{Moncrief1974a,Moncrief1974b,Chandrasekhar1983}.

\begin{figure}
 \includegraphics[width=0.96\linewidth]{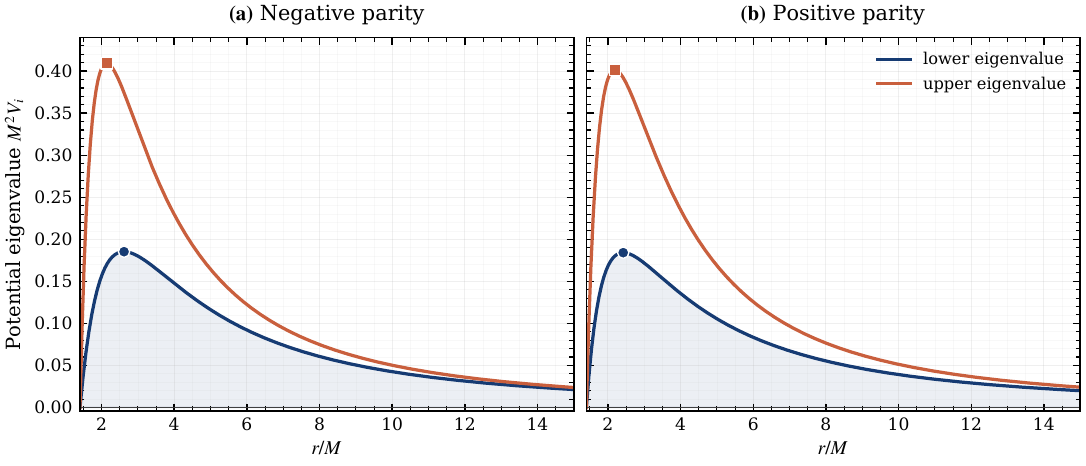}
 \caption{Canonical quadrupolar potential eigenvalues at \(\chi=0.8\): (a) negative parity and (b) positive parity. Peak markers identify the barrier radii used in the numerical analysis. Both eigenvalues remain positive on the sampled exterior grid and vanish at the horizon and at infinity.}
 \label{fig:potentials}
\end{figure}

Figure~\ref{fig:potentials} shows representative quadrupolar eigenvalues at \(\chi=0.8\).  Both eigenvalues vanish at the horizon and at infinity and remain positive over the sampled exterior domain.  Their maxima are summarized by
\begin{equation}
 M^{2}V_{\rm low}^{-}=0.18529,\quad
 M^{2}V_{\rm up}^{-}=0.40923,
\end{equation}
and
\begin{equation}
 M^{2}V_{\rm low}^{+}=0.18412,\quad
 M^{2}V_{\rm up}^{+}=0.40139.
\end{equation}

\section{Effective Reissner--Nordstr\"om correspondence}
\label{sec:rn}

Equation~\eqref{eq:asymptoticf} defines an effective charge
\begin{equation}
 \frac{Q_{\rm eff}^{2}}{M^{2}}=2\eps.
 \label{eq:qeff}
\end{equation}
We decompose the canonical operator as
\begin{equation}
 \bV_{\mathcal P}^{\rm GH}
 =
 \bV_{\mathcal P}^{\rm RN}(M,Q_{\rm eff})
 +
 \delta\bV_{\mathcal P}^{\rm NED}.
 \label{eq:decomp}
\end{equation}
This is an asymptotic reorganization, not a fit to the strong core data. It separates the leading charged geometry and Einstein Maxwell mixing from the remaining constitutive nonlinearities.

\begin{figure}
 \includegraphics[width=0.98\linewidth]{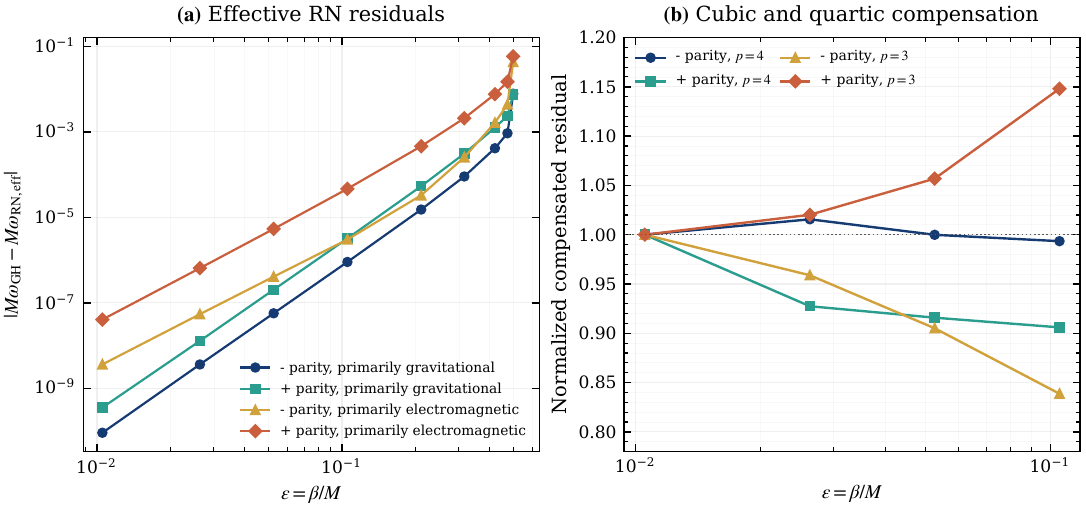}
 \caption{Weak core residual after subtraction of the effective Reissner--Nordstr\"om response. Panel (a) shows the directly computed quasinormal residuals without auxiliary guide curves. Panel (b) tests the expected cubic and quartic orders of the primarily electromagnetic and primarily gravitational branches, respectively, through compensated residuals normalized at the smallest $\eps$. Their slow variation, together with the neighboring-point slopes, supports the stated asymptotic orders without assigning unwarranted precision to a free exponent fit.}
 \label{fig:rnscaling}
\end{figure}

The background residual begins at $O(\eps^{4})$ by Eq.~\eqref{eq:asymptoticf}, whereas the electromagnetic principal coefficients contain lower order constitutive terms. The expected hierarchy is therefore
\begin{equation}
 \delta\omega_{g}^{\mathcal P}=O(\eps^{4}),
 \qquad
 \delta\omega_{e}^{\mathcal P}=O(\eps^{3}).
 \label{eq:rnpowers}
\end{equation}
We test these orders without treating a sparsely sampled exponent as a high precision fitted quantity. For the first four weak core points, the local logarithmic slopes lie in the intervals $3.92$--$4.02$ for the primarily gravitational branches and $2.89$--$3.12$ for the primarily electromagnetic branches. The corresponding compensated residuals $\delta\omega_g/\eps^4$ and $\delta\omega_e/\eps^3$ remain slowly varying over the same interval. Independent conversion and bright-channel residuals show the same quartic leading behavior in the real frequency sector. Figure~\ref{fig:rnscaling} and Table~\ref{tab:rn_scaling} report these diagnostics.

\begin{table}
\caption{Weak core diagnostics for the residual relative to the effective Reissner--Nordstr\"om operator. Local slopes are computed between the first four sampled points; the compensated interval is $\delta\omega/\epsilon^p$.}
\label{tab:rn_scaling}
\begin{ruledtabular}
\begin{tabular}{ccccc}
parity & branch & $p$ & local slope range & compensated interval\\
\hline
$-$ & primarily gravitational & 4 & 3.98--4.02 & $0.00737$--$0.00753$ \\
$-$ & primarily electromagnetic & 3 & 2.89--2.95 & $0.0026$--$0.0031$ \\
$+$ & primarily gravitational & 4 & 3.92--3.98 & $0.0262$--$0.0289$ \\
$+$ & primarily electromagnetic & 3 & 3.02--3.12 & $0.0348$--$0.04$ \\
\end{tabular}
\end{ruledtabular}
\end{table}

\section{Coupled quasinormal spectrum}
\label{sec:qnm}

We impose ingoing behavior at the event horizon and outgoing behavior at infinity,
\begin{equation}
 \bPsi_{\mathcal P}\sim
 \begin{cases}
 e^{-i\omega r_{*}}\bm a_{\rm H},&r_{*}\to-\infty,\\
 e^{+i\omega r_{*}}\bm a_{\infty},&r_{*}\to+\infty.
 \end{cases}
 \label{eq:qnbc}
\end{equation}
The boundary factors are removed analytically and the remaining functions are represented on a compact radial interval, following standard continued fraction and pseudospectral formulations of black hole quasinormal boundary value problems \cite{Leaver1985,Jansen2017,BertiCardosoStarinets2009}.  Modes are accepted only when they are stable under resolution changes, have small polynomial residuals, and continue smoothly in \(\chi\).  The channel label is assigned by the asymptotic canonical fraction and followed continuously through the parameter scan.

\begin{figure}
 \includegraphics[width=0.98\linewidth]{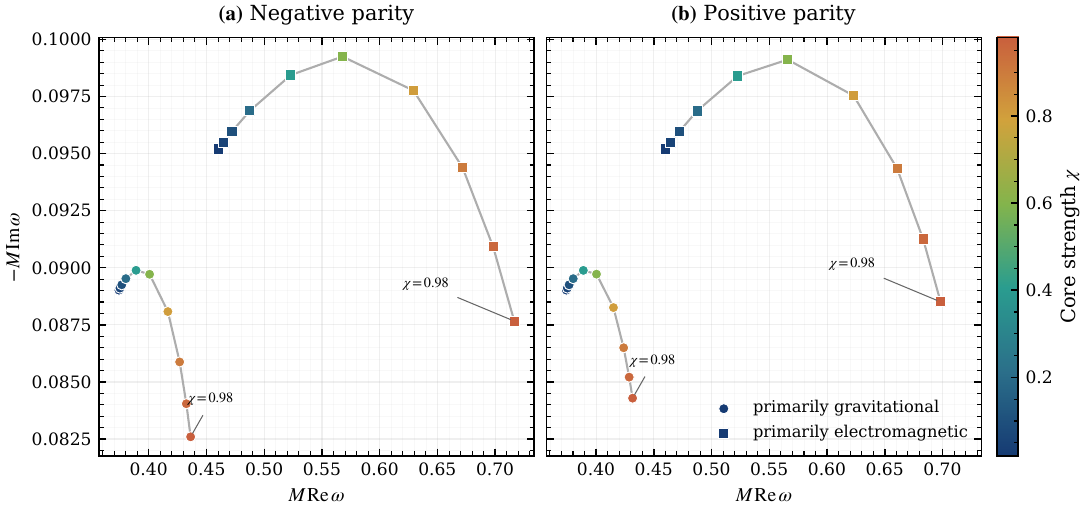}
 \caption{Fundamental quadrupolar quasinormal trajectories for (a) negative and (b) positive parity. Markers denote the sampled core strengths and the annotations indicate increasing \(\chi\). The lower and upper trajectories are the primarily gravitational and primarily electromagnetic branches, identified by continuous eigenvector overlap and asymptotic channel fractions.}
 \label{fig:qnm}
\end{figure}

Figure~\ref{fig:qnm} displays the fundamental trajectories.  At \(\chi=0.8\),
\begin{align}
 M\omega_{g}^{-}&=0.416583-0.088081i,\nonumber\\
 M\omega_{e}^{-}&=0.629221-0.097760i,\nonumber\\
 M\omega_{g}^{+}&=0.415036-0.088256i,\nonumber\\
 M\omega_{e}^{+}&=0.623005-0.097533i.
 \label{eq:qnm08}
\end{align}
At \(\chi=0.98\),
\begin{align}
 M\omega_{g}^{-}&=0.436267-0.082598i,\nonumber\\
 M\omega_{e}^{-}&=0.716507-0.087663i,\nonumber\\
 M\omega_{g}^{+}&=0.431602-0.084289i,\nonumber\\
 M\omega_{e}^{+}&=0.698262-0.088507i.
 \label{eq:qnm098}
\end{align}
The complete fundamental and first overtone set is retained in the branch tracking and convergence analysis; representative values are reported below. We call the two continuously tracked solutions the primarily gravitational and primarily electromagnetic branches according to their asymptotic canonical channel fractions; the labels \(g\) and \(e\) below refer to this dominance and not to exactly decoupled fields.

\begin{table}
\caption{Fundamental quadrupolar quasinormal modes.  The channel fraction is the squared norm fraction of the first, gravitational canonical component; the complementary fraction equals one minus the listed value.}
\label{tab:qnm}
\begin{ruledtabular}
\begin{tabular}{cccccc}
$\chi$ & parity & branch & $M\omega$ & first channel fraction & resolution difference\\
\hline
0.80 & $-$ & g & $0.416583-0.088081i$ & 0.8177 & 8.6e-12 \\
0.80 & $-$ & em & $0.629221-0.097760i$ & 0.1817 & 5.2e-12 \\
0.98 & $-$ & g & $0.436267-0.082598i$ & 0.7996 & 3.8e-11 \\
0.98 & $-$ & em & $0.716507-0.087663i$ & 0.1992 & 2.3e-11 \\
0.80 & $+$ & g & $0.415036-0.088256i$ & 0.8079 & 1.9e-11 \\
0.80 & $+$ & em & $0.623005-0.097533i$ & 0.1920 & 1.3e-11 \\
0.98 & $+$ & g & $0.431602-0.084289i$ & 0.7742 & 4.0e-11 \\
0.98 & $+$ & em & $0.698262-0.088507i$ & 0.2229 & 2.4e-11 \\
\end{tabular}
\end{ruledtabular}
\end{table}

A useful intrinsic resolvability measure is
\begin{equation}
 \mathfrak S_{\mathcal P}
 =
 \frac{|\Re\omega_{e}^{\mathcal P}-\Re\omega_{g}^{\mathcal P}|}
 {|\Im\omega_{e}^{\mathcal P}|+|\Im\omega_{g}^{\mathcal P}|}.
 \label{eq:sepindex}
\end{equation}
The branch pair becomes linewidth separated when \(\mathfrak S_{\mathcal P}>1\). Linear interpolation between the neighboring samples at \(\chi=0.6\) and \(0.8\) gives the approximate crossings
\begin{equation}
 \chi_{\rm sep}^{-}\simeq0.70,\qquad
 \chi_{\rm sep}^{+}\simeq0.71.
 \label{eq:thresholds}
\end{equation}
Only two decimal places are retained because the threshold is interpolation derived rather than directly resolved on a refined \(\chi\) grid.
The parity difference is smaller.  At \(\chi=0.98\), the relative parity asymmetries are \(1.12\times10^{-2}\) for the primarily gravitational branch and \(2.56\times10^{-2}\) for the primarily electromagnetic branch.  Thus the branch doublet is the leading spectroscopic feature, while parity splitting is a secondary consistency observable.
For later comparison we define the mass independent branch ratios and parity asymmetries by
\begin{align}
 \mathcal R_R^{\mathcal P}&=\frac{\Re\omega_e^{\mathcal P}}{\Re\omega_g^{\mathcal P}},
 &\mathcal R_I^{\mathcal P}&=\frac{|\Im\omega_e^{\mathcal P}|}{|\Im\omega_g^{\mathcal P}|},\\
 \mathcal A_j&=\frac{|\omega_j^{+}-\omega_j^{-}|}
 {\tfrac12(|\omega_j^{+}|+|\omega_j^{-}|)},
 &j&=g,e.
 \label{eq:massindependentdefs}
\end{align}

\begin{figure}
 \includegraphics[width=0.92\linewidth]{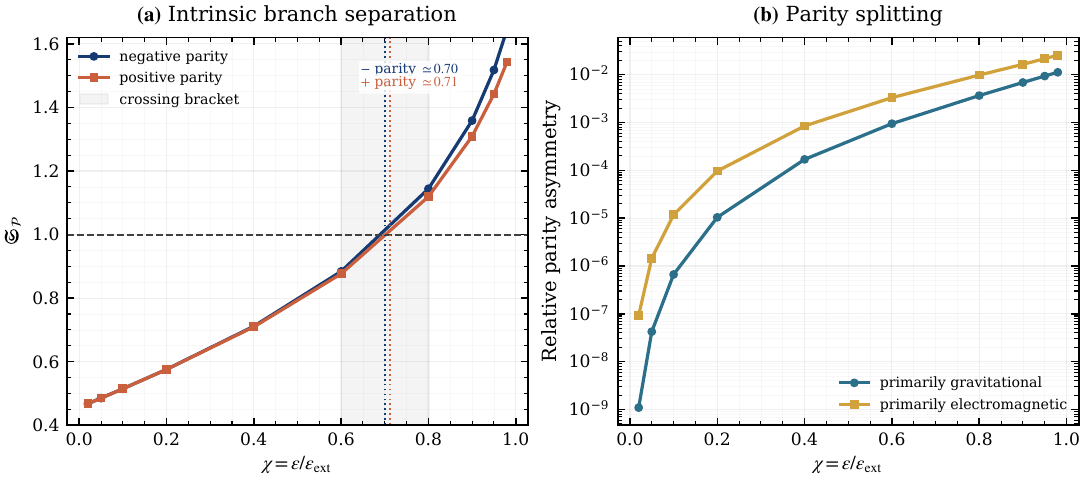}
 \caption{Intrinsic spectral identifiability. Panel (a) shows the branch separation normalized by the combined linewidth; the horizontal line marks the separation criterion, the shaded interval identifies the bracketing samples, and the vertical dotted lines show the approximate interpolated crossings. Panel (b) shows the relative odd--even parity asymmetry for the two continuously tracked branches.}
 \label{fig:linewidth}
\end{figure}

\begin{table}
\caption{Mass independent spectroscopy ratios and parity asymmetries.}
\label{tab:massindependent}
\begin{ruledtabular}
\begin{tabular}{ccccccc}
$\chi$ & $\mathcal R_R^-$ & $\mathcal R_R^+$ & $\mathcal R_I^-$ & $\mathcal R_I^+$ & $\mathcal A_g$ & $\mathcal A_{\rm em}$\\
\hline
0.80 & 1.510434 & 1.501087 & 1.109881 & 1.105113 & 3.663e-03 & 9.816e-03 \\
0.90 & 1.573899 & 1.558566 & 1.099142 & 1.090722 & 6.871e-03 & 1.648e-02 \\
0.95 & 1.613973 & 1.593735 & 1.081813 & 1.070989 & 9.369e-03 & 2.159e-02 \\
0.98 & 1.642359 & 1.617839 & 1.061331 & 1.050037 & 1.123e-02 & 2.562e-02 \\
\end{tabular}
\end{ruledtabular}
\end{table}

Time domain evolutions provide an independent check of the fundamental poles using characteristic integration and multi component ringdown fits \cite{GundlachPricePullin1994}.  Joint fits to both canonical components identify the same two branches and agree with the frequency domain values to absolute complex differences between \(4.9\times10^{-3}\) and \(1.0\times10^{-2}\) at \(\chi=0.8\), with normalized waveform residuals close to \(4\times10^{-3}\).  The comparison is used as an independent branch identification and consistency check rather than as a precision estimator because the two damped components overlap over the finite fitting window.

\section{Metric and optical characteristics}
\label{sec:characteristics}

The metric characteristic potential is
\begin{equation}
 W_{g}(r)=\frac{f(r)}{r^{2}},
 \label{eq:Wg}
\end{equation}
whereas the nonlinear electromagnetic characteristic potential is
\begin{equation}
 W_{o}(r)=\frac{f(r)\vk(r)}{r^{2}}.
 \label{eq:Wo}
\end{equation}
The characteristic radii satisfy
\begin{equation}
 W_{a}'(r_{a})=0,\qquad a=g,o.
 \label{eq:charcond}
\end{equation}
The associated orbital frequencies and impact parameters are
\begin{equation}
 \Omega_{a}=\sqrt{W_{a}(r_{a})},
 \qquad
 b_{a}=\Omega_{a}^{-1},
 \label{eq:omegaimpact}
\end{equation}
and the Lyapunov exponents follow from the second derivative of \(W_a\) with respect to the appropriate tortoise coordinate.

The metric and optical spheres are nearly degenerate at small \(\eps\), but separate monotonically toward extremality. This distinction is the nonlinear electrodynamic extension of the eikonal relation between unstable characteristic orbits and high multipole quasinormal spectra \cite{CardosoEtAl2009,ToshmatovEtAl2018,ToshmatovEtAl2018Polar}.  At \(\chi=0.8\),
\begin{equation}
 \frac{r_g}{M}=2.257002,\qquad
 \frac{r_o}{M}=2.288819,
 \label{eq:rchar08}
\end{equation}
while at \(\chi=0.98\),
\begin{equation}
 \frac{r_g}{M}=1.957792,\qquad
 \frac{r_o}{M}=2.050797.
 \label{eq:rchar098}
\end{equation}
The normalized critical curve splitting reaches \(3.02\%\) at \(\chi=0.98\).  Large multipole spectra reconstruct both critical families: the negative parity centroid converges to the metric critical curve, and the positive parity branches resolve the metric and optical limits.

\begin{figure}
 \includegraphics[width=0.98\linewidth]{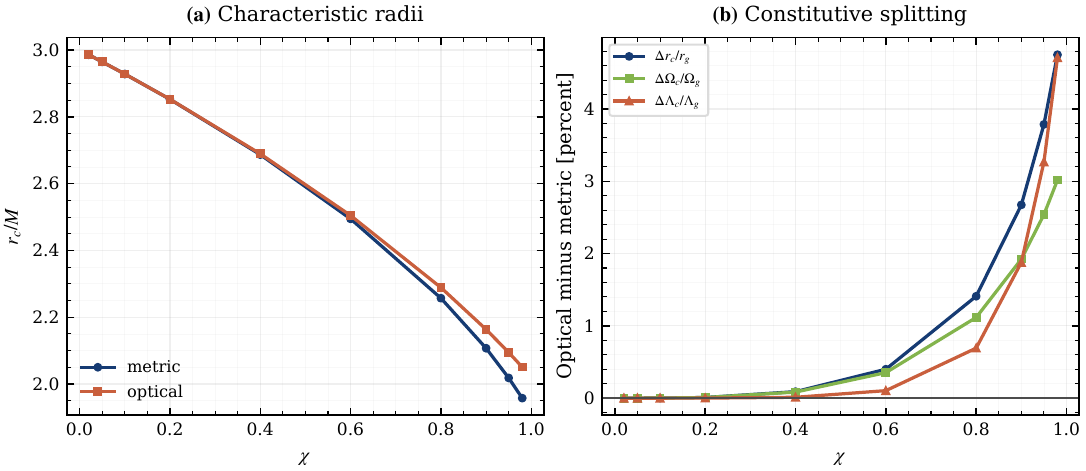}
 \caption{Separation of the metric and optical characteristic families along the black hole branch. Panel (a) shows the characteristic radii. Panel (b) gives the relative optical minus metric differences in radius, orbital frequency, and Lyapunov exponent, making explicit the distinct constitutive response that grows toward extremality.}
 \label{fig:charspheres}
\end{figure}

\begin{table}
\caption{Metric and optical characteristic spheres.}
\label{tab:characteristics}
\begin{ruledtabular}
\begin{tabular}{cccccc}
$\chi$ & family & $r_c/M$ & $M\Omega_c$ & $M\Lambda_c$ & $b_c/M$\\
\hline
0.80 & metric & 2.257002 & 0.233917 & 0.192112 & 4.275019 \\
0.80 & optical & 2.288818 & 0.231321 & 0.193441 & 4.322998 \\
0.90 & metric & 2.107215 & 0.243517 & 0.186192 & 4.106489 \\
0.90 & optical & 2.163551 & 0.238879 & 0.189687 & 4.186226 \\
0.98 & metric & 1.957792 & 0.253240 & 0.176228 & 3.948826 \\
0.98 & optical & 2.050797 & 0.245715 & 0.184528 & 4.069758 \\
\end{tabular}
\end{ruledtabular}
\end{table}

\section{Extremal throat and horizon derivative hierarchy}
\label{sec:throat}

At extremality,
\begin{equation}
 f(r)=\frac{(r-r_{\rm e})^{2}}{L_{2}^{2}}+O[(r-r_{\rm e})^{3}],
 \qquad
 L_{2}^{2}=\frac{2}{f''(r_{\rm e})}.
 \label{eq:throatexp}
\end{equation}
For the present geometry,
\begin{equation}
 \frac{L_{2}}{M}=1.107878,\qquad
 \frac{r_{\rm e}}{M}=0.912622.
 \label{eq:L2}
\end{equation}
The near horizon limit is \(AdS_{2}\times S^{2}\).  Diagonalizing the reduced two channel mass matrix gives eigenvalues \(\mu_{\mathcal P j}^{2}\) and conformal weights
\begin{equation}
 h_{\mathcal P j}
 =
 \frac12+\sqrt{\frac14+\mu_{\mathcal P j}^{2}L_{2}^{2}}.
 \label{eq:h}
\end{equation}
For \(\ell=2\), the leading weights are
\begin{equation}
 h_{-,1}=2.344696,\qquad h_{+,1}=1.918073,
 \label{eq:hlead}
\end{equation}
with the second branches \(h_{-,2}=4.596486\) and \(h_{+,2}=4.133119\).

\begin{figure}
 \includegraphics[width=0.92\linewidth]{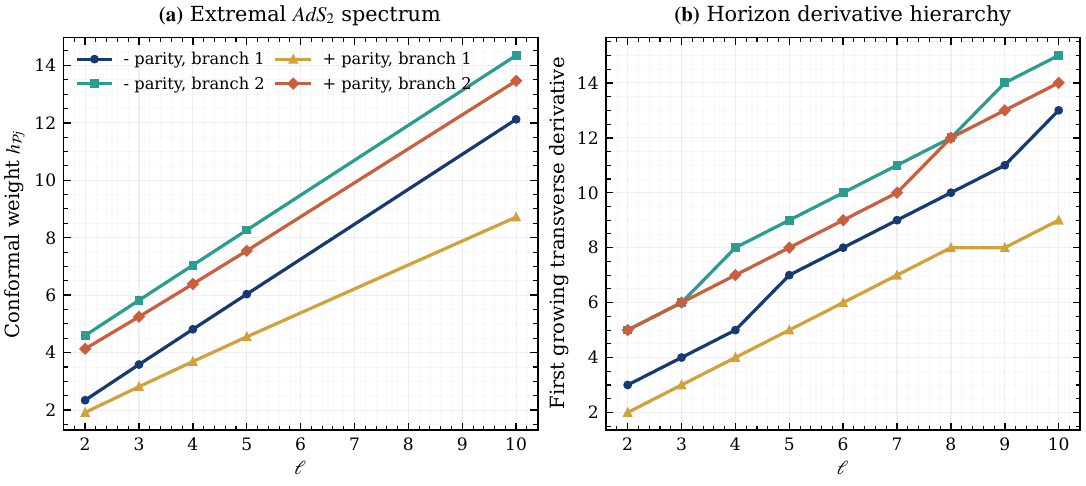}
 \caption{Extremal throat data. Panel (a) shows the two conformal weight branches in each parity sector as functions of multipole number. Panel (b) gives the first transverse derivative order expected to grow along the future horizon under the late time scaling in Eq.~\eqref{eq:aretakis}. The annotations identify the quadrupolar values used in the near extremal comparison.}
 \label{fig:weights}
\end{figure}

\begin{table}
\caption{Quadrupolar extremal throat eigenvalues and conformal weights.}
\label{tab:weights}
\begin{ruledtabular}
\begin{tabular}{ccccc}
parity & branch & $M^2\mu^2$ & $\mu^2L_2^2$ & $h$\\
\hline
$-$ & 1 & 2.568780 & 3.152904 & 2.344696 \\
$-$ & 2 & 13.468542 & 16.531200 & 4.596486 \\
$+$ & 1 & 1.434691 & 1.760930 & 1.918073 \\
$+$ & 2 & 10.550448 & 12.949551 & 4.133119 \\
\end{tabular}
\end{ruledtabular}
\end{table}

The near extremal zero damping towers satisfy
\begin{equation}
 \omega_{\mathcal P j n}
 =
 -i\kap_{+}\left(n+h_{\mathcal P j}\right)
 +O(\kap_{+}^{2}),
 \label{eq:zdm}
\end{equation}
in the static neutral frame.  At \(\chi=0.99999\), the numerical ratios
\(-\Im\omega/\kap_{+}\) differ from the predicted weights by about \(0.5\%\) for the fundamental near horizon branches.  This convergence is independently controlled by the exact extremal mass matrix.

The same conformal weights motivate an extremal horizon derivative hierarchy.  If a branch has late time behavior \(v^{-h}\) has transverse derivatives
\begin{equation}
 \left.\partial_{r}^{k}\Psi_{\mathcal P j}\right|_{\mathcal H^{+}}
 \sim v^{\,k-h_{\mathcal P j}},
 \label{eq:aretakis}
\end{equation}
up to branch dependent amplitudes.  The first derivative expected to grow is then the smallest integer \(k>h_{\mathcal P j}\).  This provides a direct consistency link between near horizon damping and the associated Aretakis type hierarchy \cite{Aretakis2011,LuciettiReall2012}.

\section{Quadrupolar real frequency scattering and coherent absorption}
\label{sec:scattering}

The quantitative scattering, packet, and coherent-channel results in this section are restricted to \(\ell=2\). This sector contains the lowest radiative multipole and is used consistently for all cross comparisons below.

\subsection{Canonical scattering matrix}

For real \(\omega>0\), a unit incoming amplitude vector at infinity defines
\begin{align}
 \bPsi_{\mathcal P}
 &\sim e^{-i\omega r_*}\bm a_{\rm in}
 +e^{+i\omega r_*}\bR_{\mathcal P}\bm a_{\rm in},
 &&r_*\to+\infty,
 \label{eq:asympinfty}\\
 \bPsi_{\mathcal P}
 &\sim e^{-i\omega r_*}\bT_{\mathcal P}\bm a_{\rm in},
 &&r_*\to-\infty.
 \label{eq:asymphorizon}
\end{align}
Wronskian conservation gives
\begin{equation}
 \bR_{\mathcal P}^{\dagger}\bR_{\mathcal P}
 +
 \bT_{\mathcal P}^{\dagger}\bT_{\mathcal P}
 =\id.
 \label{eq:unitarity}
\end{equation}
Reciprocity in the real canonical basis implies a symmetric scattering response after consistent phase fixing.  Across the full production grid, the maximum direct flux-balance and pointwise unitarity residuals are \(2.43\times10^{-8}\); at the conversion peaks they decrease below \(2\times10^{-11}\). Peak locations are obtained from dedicated locally refined frequency scans rather than from the presentation grid. Dedicated calculations with extraction radii between \(140M\) and \(350M\) change the peak conversion probability by at most \(2.41\times10^{-4}\).

For an incident gravitational channel, the reflected conversion probability is
\begin{equation}
 \mathscr C_{\mathcal P}(\omega)
 =
 \left|\left(\bR_{\mathcal P}\right)_{eg}\right|^{2}.
 \label{eq:conversion}
\end{equation}
The corresponding total gravitational-to-electromagnetic transfer, including the converted flux reflected to infinity and transmitted through the horizon, is
\begin{equation}
 \mathscr X_{\mathcal P}(\omega)
 =\left|\left(\bR_{\mathcal P}\right)_{eg}\right|^{2}
 +\left|\left(\bT_{\mathcal P}\right)_{eg}\right|^{2}.
 \label{eq:totaltransfer}
\end{equation}
At \(\chi=0.9\),
\begin{equation}
 \mathscr C_{-}^{\rm max}\simeq0.3247,\qquad
 \mathscr C_{+}^{\rm max}\simeq0.4576.
 \label{eq:conversionmax}
\end{equation}
The corresponding refined peak frequencies are \(M\omega\simeq0.43038\) and \(0.40563\).  The effective Reissner--Nordstr\"om response captures most of the conversion, while the nonlinear residual is below one percent at these peaks.

\begin{figure}
 \includegraphics[width=0.98\linewidth]{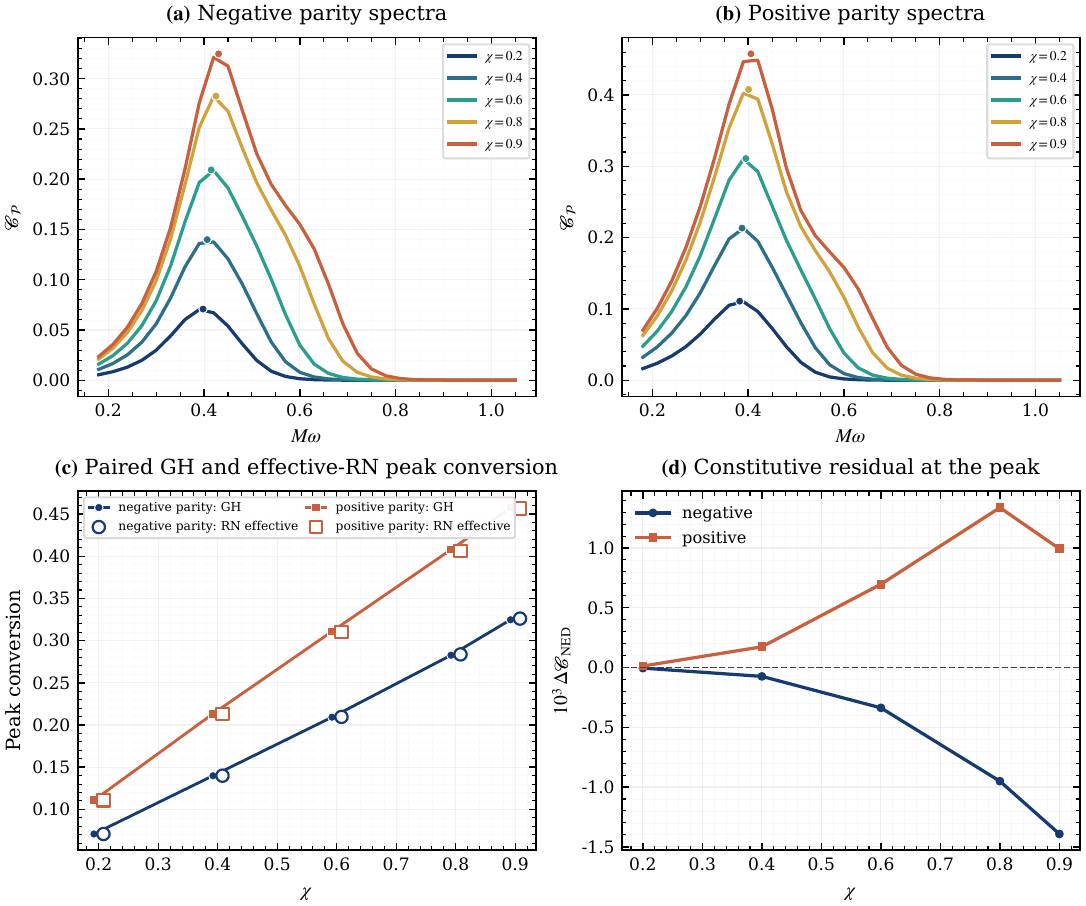}
 \caption{Reflected gravitoelectromagnetic conversion. Panels (a) and (b) show the negative and positive parity quadrupolar spectra for five core strengths, with color encoding \(\chi\) and peak markers separating nearby curves. Panel (c) compares the generalized Hayward peak conversion with the effective Reissner--Nordstr\"om values evaluated at the same peak frequencies. Filled markers joined by solid curves denote the generalized Hayward result, while larger open markers at the same \(\chi\) values denote the effective charged comparator. Panel (d) reports the unshifted signed difference. The effective charged operator captures the dominant conversion, while the remaining subpercent residual isolates the nonlinear constitutive correction.}
 \label{fig:conversion}
\end{figure}

\begin{table}
\caption{Peak reflected conversion and total cross transfer for the quadrupolar sector. Peak locations are obtained from dedicated refined scans; the last column reports pointwise unitarity closure.}
\label{tab:conversion}
\begin{ruledtabular}
\begin{tabular}{cccccc}
$\chi$ & parity & $M\omega_{\rm pk}$ & $\mathscr C_{\rm pk}$ & $\mathscr X_{\rm pk}$ & unitarity error\\
\hline
0.8 & $+$ & 0.40071 & 0.4078 & 0.4689 & $6.6e-12$ \\
0.8 & $-$ & 0.42484 & 0.2827 & 0.3712 & $1.7e-11$ \\
0.9 & $+$ & 0.40563 & 0.4576 & 0.5184 & $1.0e-12$ \\
0.9 & $-$ & 0.43038 & 0.3247 & 0.4070 & $2.0e-12$ \\
\end{tabular}
\end{ruledtabular}
\end{table}

\subsection{Finite bandwidth transfer}

For a normalized incident packet with spectral amplitude \(A(\omega)\), the outgoing conversion efficiency is
\begin{equation}
 \overline{\mathscr C}_{\mathcal P}
 =
 \int_{0}^{\infty}d\omega\,|A(\omega)|^{2}
 \mathscr C_{\mathcal P}(\omega).
 \label{eq:packet}
\end{equation}
For a packet family of fixed width \(\sigma_\omega\), we denote the efficiency optimized over its central frequency by
\begin{equation}
 \eta_{\mathcal P}(\sigma_\omega)
 =\max_{\omega_0}\overline{\mathscr C}_{\mathcal P}(\omega_0,\sigma_\omega).
 \label{eq:eta}
\end{equation}
At \(\chi=0.9\) and positive parity, a narrow packet with \(M\sigma_{\omega}=0.01\) reaches \(45.3\%\), while \(M\sigma_{\omega}=0.04\) retains \(40.8\%\).  The same widths give \(32.1\%\) and \(28.9\%\) in negative parity.  The persistence under bandwidth averaging shows that the conversion peak is not solely a monochromatic feature of the quadrupolar response.

\begin{figure}
 \includegraphics[width=0.92\linewidth]{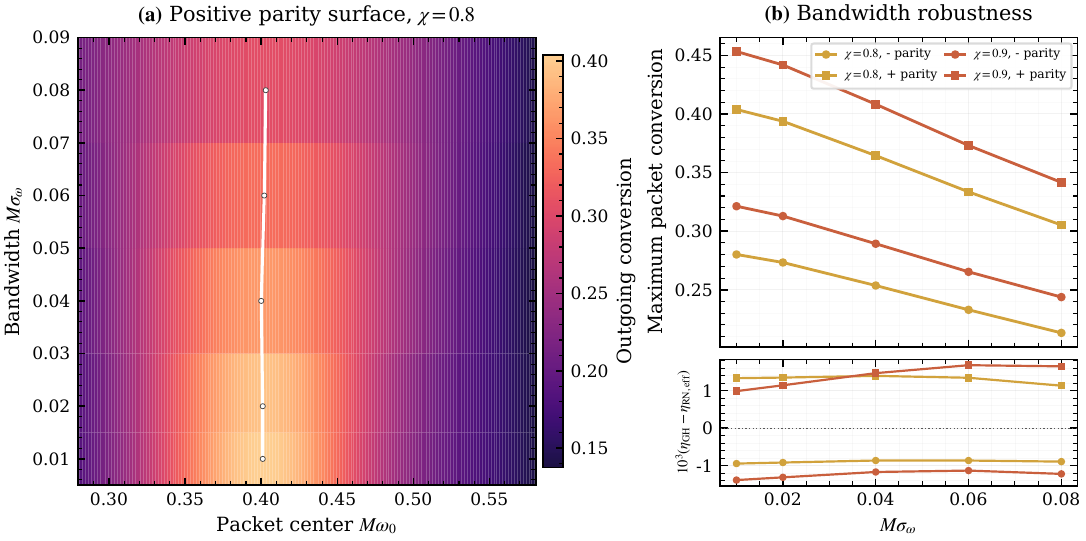}
 \caption{Finite bandwidth conversion. Panel (a) shows the positive parity conversion surface at \(\chi=0.8\), with the white ridge tracing the optimal center frequency. The upper part of panel (b) shows the generalized Hayward maximum packet conversion versus bandwidth for both parities at \(\chi=0.8\) and \(0.9\). The lower residual panel gives \(10^3(\eta_{\rm GH}-\eta_{\rm RN,eff})\), replacing nearly indistinguishable overplotted baselines by a directly resolved constitutive difference.}
 \label{fig:packet}
\end{figure}

\subsection{Bright and dark greybody eigenchannels}

The absorption operator is
\begin{equation}
 \bA_{\mathcal P}
 =
 \id-\bR_{\mathcal P}^{\dagger}\bR_{\mathcal P}
 =
 \bT_{\mathcal P}^{\dagger}\bT_{\mathcal P}.
 \label{eq:absorption}
\end{equation}
It is Hermitian and contractive.  Its eigenvalues
\begin{equation}
 0\leq\Gamma_{\rm d}\leq\Gamma_{\rm b}\leq1
 \label{eq:eigenvalues}
\end{equation}
define dark and bright coherent incident channels.  The eigenvectors specify the relative gravitational and electromagnetic amplitude and phase required for extremal absorption.

At \(\chi=0.9\), \(\ell=2\), and \(M\omega=0.545714\), we obtain
\begin{align}
 (\Gamma_{\rm d}^{-},\Gamma_{\rm b}^{-})
 &=(0.012736,0.983463),\\
 (\Gamma_{\rm d}^{+},\Gamma_{\rm b}^{+})
 &=(0.018527,0.984437).
 \label{eq:brightdark}
\end{align}
Writing a normalized bright eigenvector as
\begin{equation}
 \bm a_{\rm b}=
 \begin{pmatrix}
 \cos\theta_{\rm b}\\ e^{i\phi_{\rm b}}\sin\theta_{\rm b}
 \end{pmatrix},
 \label{eq:brightangle}
\end{equation}
the canonical electromagnetic amplitude angle is near \(26^\circ\); the quoted eigenvalues refer to this common comparison frequency, not to the conversion-peak frequencies.  The large contrast is a coherent property of the matrix response and cannot be inferred from either pure channel absorption probability separately.

\begin{figure}
 \includegraphics[width=0.98\linewidth]{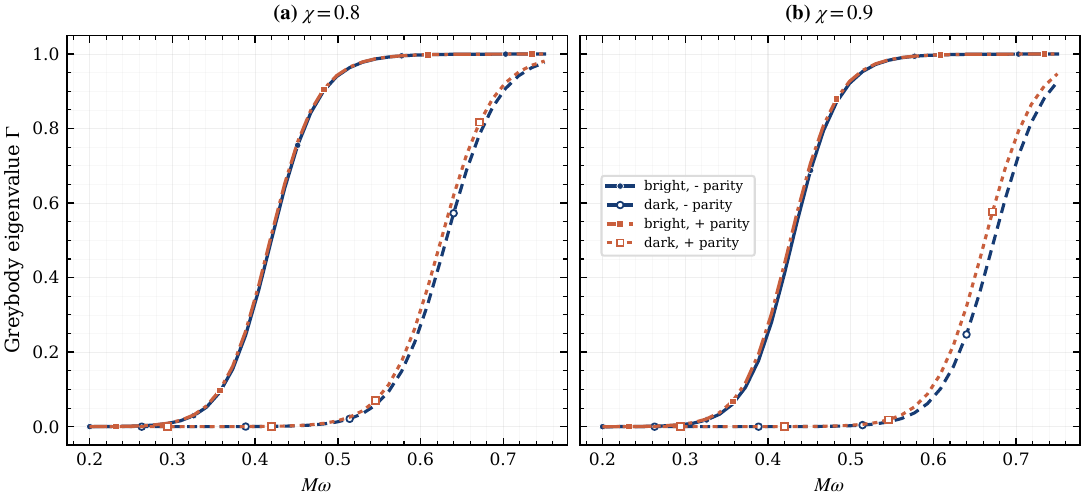}
 \caption{Dark and bright greybody eigenvalues for (a) \(\chi=0.8\) and (b) \(\chi=0.9\), with both parity sectors shown. Solid curves with filled sparse markers denote the bright eigenchannels, while dashed curves with open staggered markers denote the dark eigenchannels; color and marker shape distinguish parity. The sparse markers expose the small parity offsets where the curves nearly coincide without obscuring the continuous frequency dependence. Their large eigenvalue gap across the barrier region corresponds to near maximal and near minimal horizon coupling.}
 \label{fig:greybody}
\end{figure}

\begin{table}
\caption{Dark and bright quadrupolar absorption eigenchannels at $\chi=0.9$ and the common comparison frequency $M\omega=0.545714$.}
\label{tab:coherent}
\begin{ruledtabular}
\begin{tabular}{cccccc}
parity & $M\omega$ & $\Gamma_{\rm d}$ & $\Gamma_{\rm b}$ & $\theta_{\rm b}$ (deg) & closure error\\
\hline
$-$ & 0.545714 & 0.0127 & 0.9835 & 26.02 & $7.0e-12$ \\
$+$ & 0.545714 & 0.0185 & 0.9844 & 26.58 & $7.2e-12$ \\
\end{tabular}
\end{ruledtabular}
\end{table}

\section{Cross observable consistency}
\label{sec:closure}

The independently evaluated sectors provide four internal checks.

First, the asymptotic geometry fixes $Q_{\rm eff}^{2}/M^{2}=2\eps$. The same effective charge captures the leading quasinormal shifts, conversion spectrum, and bright greybody response. Local slopes and compensated residuals support cubic spectral corrections in the primarily electromagnetic branch and quartic residuals in the primarily gravitational branch and real frequency response.

Second, the constitutive factor $\vk$ controls both exterior hyperbolicity and the optical characteristic potential. At extremality the independently evaluated throat slopes satisfy
\begin{equation}
 \left(\frac{\alpha_{o}}{\alpha_{g}}\right)^{2}
 =
 \vk_{\rm e},
 \label{eq:opticalclosure}
\end{equation}
with absolute residual $6.24\times10^{-13}$.

Third, the extremal reduced mass matrix determines the near horizon damping ratios and the horizon derivative hierarchy through the same conformal weights, without an additional fitted parameter.

Fourth, the canonical normalization used for quasinormal modes also yields a unitary real frequency scattering matrix. The Wronskian closure and the agreement between $\id-\bR^\dagger\bR$ and $\bT^\dagger\bT$ independently audit the real frequency implementation.

\section{Discussion}
\label{sec:discussion}

The dominant weak core response is effectively charged rather than intrinsically nonlinear. The asymptotic Reissner--Nordstr\"om operator accounts for most of the gravitoelectromagnetic mixing and conversion. A large conversion probability is therefore not by itself a distinctive nonlinear electrodynamic signature. Constitutive information resides in the compensated residual hierarchy, parity asymmetry, metric--optical characteristic splitting, and agreement among observables derived from the same operator.

The largest intrinsic spectral effect is the separation between primarily gravitational and primarily electromagnetic branches. In the quadrupolar sector it exceeds the combined linewidth at $\chi\simeq0.70$, whereas the parity difference remains smaller. The branch doublet is consequently the primary intrinsic spectroscopic feature of this model, with parity splitting serving as a secondary consistency observable.

Pure gravitational or electromagnetic incidence does not diagonalize horizon absorption. The bright and dark eigenvectors vary slowly across the analyzed resonance band, so a fixed coherent preparation remains meaningful for finite bandwidth packets. This matrix structure is independent of any detector model and is a property of the canonical exterior response.

Because the equations are scale free, a dimensionless mode maps kinematically to the observed frequency through $f_{\rm obs}=\operatorname{Re}(M\omega)/(2\pi M_z)$, with $M_z=(1+z)M$. We do not attach a detectability claim to this conversion, since mode excitation, source overlap, detector response, and inference systematics must be modeled separately \cite{BertiCardosoWill2006,SilvaEtAl2024,BertiEtAl2026}.

No exterior linear instability is found within the investigated parity sectors, multipole range, parameter domain, and spectral window. The exterior satisfies $\Lag_{\F}>0$, $\vk>0$, regular canonical transformations, positive sampled potential eigenvalues, and decaying computed modes. These checks establish a regular and numerically consistent exterior wave problem for the chosen branch; they are not statements about the central region or nonlinear evolution of the complete spacetime.

\section{Conclusions}
\label{sec:conclusion}

We constructed the coupled gravitational and electromagnetic response of a magnetically supported generalized Hayward black hole from a specified nonlinear electrodynamic completion. The reconstruction yields a Maxwell weak field limit, the effective charge relation $Q_{\rm eff}^{2}/M^{2}=2\eps$, and the exterior optical factor $\vk=(2-5y)/[2(1+y)]$.

Both parity sectors reduce to real symmetric two channel wave equations. After subtraction of the effective Reissner--Nordstr\"om contribution, local slopes and compensated residuals are consistent with cubic spectral corrections in the primarily electromagnetic branch and quartic corrections in the primarily gravitational branch and principal scattering observables.

The quadrupolar ringdown contains primarily gravitational and primarily electromagnetic branches whose intrinsic separation exceeds their combined linewidth at $\chi\simeq0.70$. Parity splitting remains smaller but grows toward extremality. The short wavelength limit resolves distinct metric and optical characteristic spheres, with a critical curve difference of about $3\%$ at $\chi=0.98$.

The extremal throat reduction gives coupled $AdS_{2}$ masses and conformal weights that reproduce the near extremal damping ratios and determine the onset order of growing transverse horizon derivatives. In the real frequency quadrupolar sector, positive parity reflected conversion reaches $45.8\%$ at $\chi=0.9$ and remains above $40\%$ after moderate bandwidth averaging. Diagonalization of the absorption matrix yields simultaneous bright absorption above $98\%$ and dark absorption below $2\%$.

The quasinormal spectrum, characteristic structure, conversion matrix, packet response, and absorption eigenchannels thus constitute a coherent set of diagnostics for one nonlinear electrodynamic operator. Their common effective charged limit and independently verified closure relations provide the central internal consistency test of the analysis.

\section*{Data Availability}
This article contains all data generated or analyzed during the current study; no additional datasets are available.

\appendix

\section{Canonical reconstruction and operator checks}
\label{app:operator}

The invariant convention is \(\F=F_{\mu\nu}F^{\mu\nu}/4\), with Maxwell limit \(\Lag=\F\), magnetic invariant \(Q_{\rm m}^{2}/(2r^{4})\), and action density \(R-4\Lag\). The background reconstruction can be verified directly in radial form,
\begin{equation}
 \Lag(r)=\frac{m'(r)}{r^{2}}
 =\frac{M\beta}{(r^{3}+\beta^{3})^{4/3}},
\end{equation}
which, together with \(Q_{\rm m}^{2}=2M\beta\), reproduces Eq.~\eqref{eq:L} and identically satisfies the angular Einstein equation.

After reducing the gauge invariant perturbations to a quadratic action, a radial field redefinition removes first derivative mixing and normalizes the kinetic matrix to the identity. The remaining constant asymptotic rotation is fixed continuously from the Schwarzschild limit so that the two canonical components carry unit gravitational and electromagnetic flux. With
\begin{equation}
 L=\ell(\ell+1),\qquad \lambda=(\ell-1)(\ell+2)=L-2,
\end{equation}
and
\begin{align}
 \mathcal D_{-}&=\Lag_{\F}^{1/2}\frac{d}{dr}
 \left[f\frac{d}{dr}\Lag_{\F}^{-1/2}\right],\\
 \mathcal D_{+}&=\Lag_{\F}^{-1/2}\frac{d}{dr}
 \left[f\frac{d}{dr}\Lag_{\F}^{1/2}\right],
 \label{eq:Dpm}
\end{align}
the odd parity entries are
\begin{align}
 U_{-}^{gg}&=\frac{L}{r^{2}}-\frac{6m}{r^{3}}+2\Lag,\\
 U_{-}^{ge}&=+\frac{2Q_{\rm m}\sqrt{\lambda\Lag_{\F}}}{r^{3}},\\
 U_{-}^{ee}&=\frac{L}{r^{2}}+\frac{4Q_{\rm m}^{2}\Lag_{\F}}{r^{4}}+\mathcal D_{-}.
 \label{eq:Uminus}
\end{align}
For even parity, define
\begin{align}
 a&=\frac{6m}{r}-2r^{2}\Lag,
 &b&=\lambda+\frac{4Q_{\rm m}^{2}\Lag_{\F}}{r^{2}},\\
 d&=a+\lambda,
 &c_{1}&=\lambda+1-f+2r^{2}\Lag,\\
 c_{2}&=c_{1}+4f\vk,
 &w&=c_{1}+2f\vk.
 \label{eq:evenaux}
\end{align}
The canonical entries are
\begin{align}
 U_{+}^{gg}&=
 \frac{L\lambda-2f\lambda+a(a-4m/r)}{r^{2}d}
 +\frac{2f\lambda b}{r^{2}d^{2}},
 \label{eq:Uplusgg}\\
 U_{+}^{ee}&=
 \frac{\vk L}{r^{2}}+\mathcal D_{+}
 +\frac{4Q_{\rm m}^{2}\Lag_{\F}c_{2}}{r^{4}d}
 +\frac{8fQ_{\rm m}^{2}\Lag_{\F}b}{r^{4}d^{2}},
 \label{eq:Uplusee}\\
 U_{+}^{ge}&=-\frac{2Q_{\rm m}\sqrt{\lambda\Lag_{\F}}}{r^{3}}
 \left(\frac{w}{d}+\frac{2fb}{d^{2}}\right).
 \label{eq:Uplusge}
\end{align}
The residual sign of either canonical electromagnetic amplitude is conventional and does not affect potential eigenvalues, quasinormal frequencies, conversion probabilities, or absorption eigenvalues. Direct substitution of the background functions reproduces the matrices used in the numerical analysis. The Maxwell limit gives the canonical Reissner--Nordstr\"om system, while \(Q_{\rm m}\to0\) yields the Regge--Wheeler and electromagnetic Schwarzschild potentials. Throughout the sampled exterior,
\begin{equation}
 \bV_{\mathcal P}^{T}=\bV_{\mathcal P},\qquad
 \bV_{\mathcal P}(r_{+})=\bm0,\qquad
 \bV_{\mathcal P}(r)=O(r^{-2}),
\end{equation}
and the even parity denominator \(d=a+\lambda\) remains positive.

\section{Numerical validation and asymptotic consistency}
\label{app:validation}

The quasinormal problem is compactified after extracting the exact ingoing and outgoing factors. Branches are tracked by eigenvector overlap and canonical channel fractions rather than by frequency ordering. The scattering fundamental matrix is integrated from a near horizon ingoing basis and matched to canonical plane waves at large radius; quoted conversion maxima are obtained from dedicated local frequency refinements. Table~\ref{tab:validation} summarizes the quantitative checks used in the production analysis.

\begin{table}[htbp]
\caption{Compact numerical validation summary.}
\label{tab:validation}
\centering
\small
\begin{tabular}{lc}
\toprule
Diagnostic & Maximum discrepancy \\
\midrule
Fundamental QNM resolution change & \(1.2\times10^{-10}\) \\
QNM polynomial residual & \(4\times10^{-16}\) \\
First overtone resolution drift & \(1.2\times10^{-8}\) \\
Full-grid flux closure residual & \(2.5\times10^{-8}\) \\
Conversion-peak closure residual & \(2.0\times10^{-11}\) \\
Extraction-radius Wronskian residual & \(4.0\times10^{-13}\) \\
Conversion shift over \(140M\le r_{\max}\le350M\) & \(2.41\times10^{-4}\) \\
Large-\(\ell\) characteristic residual & \(2.0\times10^{-3}\) \\
Near-horizon conformal-weight mismatch & \(0.54\%\) \\
\bottomrule
\end{tabular}
\end{table}

The smallest sampled potential eigenvalues are approximately \(7\times10^{-6}\) and \(8\times10^{-6}\) for the lower negative and positive parity branches, respectively, while the upper-branch minima are approximately \(2.6\times10^{-5}\) and \(2.4\times10^{-5}\). These positive sampled values complement the exterior kinetic and hyperbolicity conditions; they are numerical diagnostics on the investigated domain, not a statement about nonlinear stability.

\begin{figure}[!htbp]
 \centering
 \includegraphics[width=0.94\linewidth]{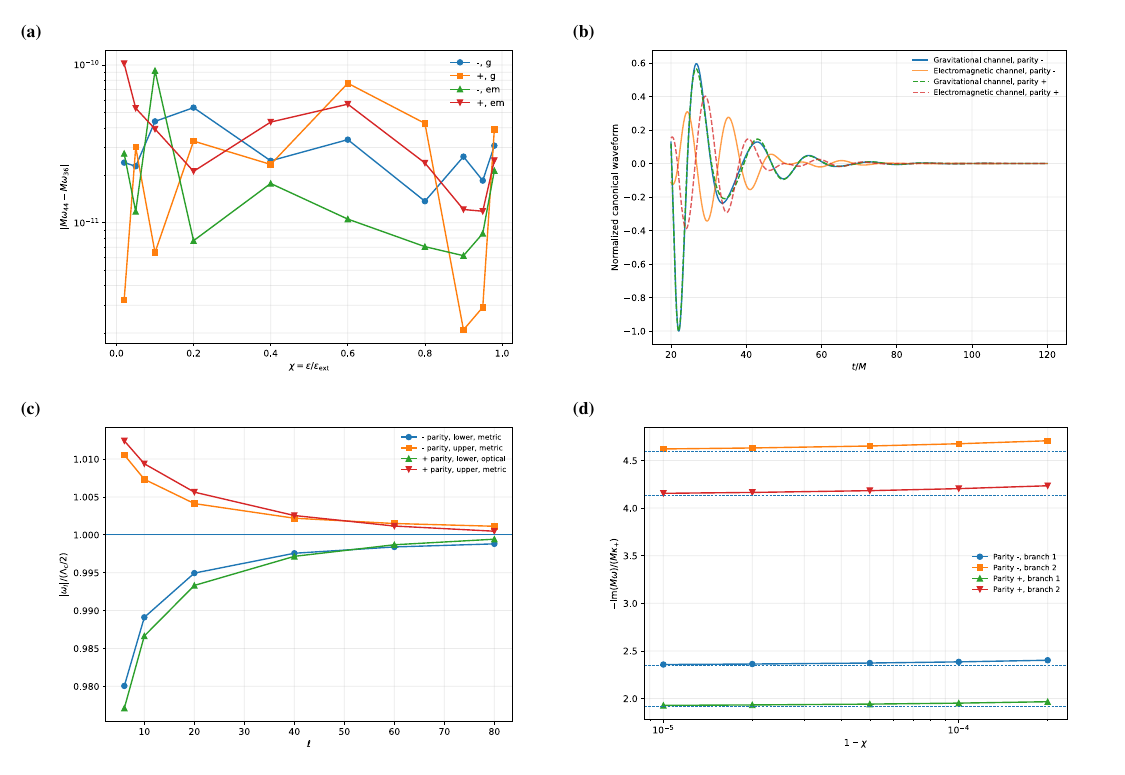}
 \caption{Compact validation suite: (a) fundamental QNM resolution convergence; (b) representative canonical time domain waveforms; (c) large-\(\ell\) convergence to metric and optical Lyapunov predictions; and (d) near-extremal convergence to the independent throat weights.}
 \label{fig:validation}
\end{figure}

Figure~\ref{fig:validation} collects the four independent checks that are most informative beyond the main figures. Panel (a) shows resolution stability of the fundamental quadrupolar modes. Panel (b) displays representative time domain evolutions, which resolve both canonical branches and provide an independent qualitative cross-check of the frequency domain spectrum. Panel (c) verifies convergence of the large-\(\ell\) damping rates to the metric and optical Lyapunov predictions. Panel (d) shows the near-extremal approach of \(-\operatorname{Im}\omega/\kappa_{+}\) to the independently evaluated throat weights.

The small-core parity splitting vanishes rapidly in the Maxwell and Schwarzschild limits, with the primarily electromagnetic branch remaining more sensitive to constitutive derivatives. After subtraction of the effective Reissner--Nordstr\"om response, the peak-conversion and bright-channel residuals approach quartic behavior, consistently with the diagnostics in Figs.~\ref{fig:rnscaling} and \ref{fig:conversion}. The finite-bandwidth trends are already displayed in Fig.~\ref{fig:packet}. Across the plotted conversion band, the bright-channel mixing angle varies by at most \(0.30^{\circ}\), supporting a fixed coherent incident preparation over a finite packet bandwidth.

The absorption eigenvalues are independently reconstructed from both \(\id-\bR^{\dagger}\bR\) and \(\bT^{\dagger}\bT\), and their spectra agree within the quoted unitarity residual. Eigenvector phases are fixed by requiring a positive gravitational component, leaving a unique relative phase modulo an overall sign.

%


\clearpage

\begin{thebibliography}{40}%
\fontsize{8.4}{9.2}\selectfont
\makeatletter
\providecommand \@ifxundefined [1]{%
 \@ifx{#1\undefined}
}%
\providecommand \@ifnum [1]{%
 \ifnum #1\expandafter \@firstoftwo
 \else \expandafter \@secondoftwo
 \fi
}%
\providecommand \@ifx [1]{%
 \ifx #1\expandafter \@firstoftwo
 \else \expandafter \@secondoftwo
 \fi
}%
\providecommand \natexlab [1]{#1}%
\providecommand \enquote  [1]{``#1''}%
\providecommand \bibnamefont  [1]{#1}%
\providecommand \bibfnamefont [1]{#1}%
\providecommand \citenamefont [1]{#1}%
\providecommand \href@noop [0]{\@secondoftwo}%
\providecommand \href [0]{\begingroup \@sanitize@url \@href}%
\providecommand \@href[1]{\@@startlink{#1}\@@href}%
\providecommand \@@href[1]{\endgroup#1\@@endlink}%
\providecommand \@sanitize@url [0]{\catcode `\\12\catcode `\$12\catcode
  `\&12\catcode `\#12\catcode `\^12\catcode `\_12\catcode `\%12\relax}%
\providecommand \@@startlink[1]{}%
\providecommand \@@endlink[0]{}%
\providecommand \url  [0]{\begingroup\@sanitize@url \@url }%
\providecommand \@url [1]{\endgroup\@href {#1}{\urlprefix }}%
\providecommand \urlprefix  [0]{URL }%
\providecommand \Eprint [0]{\href }%
\providecommand \doibase [0]{https://doi.org/}%
\providecommand \selectlanguage [0]{\@gobble}%
\providecommand \bibinfo  [0]{\@secondoftwo}%
\providecommand \bibfield  [0]{\@secondoftwo}%
\providecommand \translation [1]{[#1]}%
\providecommand \BibitemOpen [0]{}%
\providecommand \bibitemStop [0]{}%
\providecommand \bibitemNoStop [0]{.\EOS\space}%
\providecommand \EOS [0]{\spacefactor3000\relax}%
\providecommand \BibitemShut  [1]{\csname bibitem#1\endcsname}%
\let\auto@bib@innerbib\@empty
\bibitem [{\citenamefont {Chandrasekhar}(1983)}]{Chandrasekhar1983}%
  \BibitemOpen
  \bibfield  {author} {\bibinfo {author} {\bibfnamefont {S.}~\bibnamefont
  {Chandrasekhar}},\ }\href@noop {} {\emph {\bibinfo {title} {The Mathematical
  Theory of Black Holes}}}\ (\bibinfo  {publisher} {Oxford University Press},\
  \bibinfo {address} {Oxford},\ \bibinfo {year} {1983})\BibitemShut {NoStop}%
\bibitem [{\citenamefont {Berti}\ \emph {et~al.}(2009)\citenamefont {Berti},
  \citenamefont {Cardoso},\ and\ \citenamefont
  {Starinets}}]{BertiCardosoStarinets2009}%
  \BibitemOpen
  \bibfield  {author} {\bibinfo {author} {\bibfnamefont {E.}~\bibnamefont
  {Berti}}, \bibinfo {author} {\bibfnamefont {V.}~\bibnamefont {Cardoso}},\
  and\ \bibinfo {author} {\bibfnamefont {A.~O.}\ \bibnamefont {Starinets}},\
  }\bibfield  {title} {\bibinfo {title} {Quasinormal modes of black holes and
  black branes},\ }\href {https://doi.org/10.1088/0264-9381/26/16/163001}
  {\bibfield  {journal} {\bibinfo  {journal} {Class. Quantum Grav.}\ }\textbf
  {\bibinfo {volume} {26}},\ \bibinfo {pages} {163001} (\bibinfo {year}
  {2009})}\BibitemShut {NoStop}%
\bibitem [{\citenamefont {Konoplya}\ and\ \citenamefont
  {Zhidenko}(2011)}]{KonoplyaZhidenko2011}%
  \BibitemOpen
  \bibfield  {author} {\bibinfo {author} {\bibfnamefont {R.~A.}\ \bibnamefont
  {Konoplya}}\ and\ \bibinfo {author} {\bibfnamefont {A.}~\bibnamefont
  {Zhidenko}},\ }\bibfield  {title} {\bibinfo {title} {Quasinormal modes of
  black holes: From astrophysics to string theory},\ }\href
  {https://doi.org/10.1103/RevModPhys.83.793} {\bibfield  {journal} {\bibinfo
  {journal} {Rev. Mod. Phys.}\ }\textbf {\bibinfo {volume} {83}},\ \bibinfo
  {pages} {793} (\bibinfo {year} {2011})}\BibitemShut {NoStop}%
\bibitem [{\citenamefont {Berti}\ \emph {et~al.}(2026)\citenamefont {Berti},
  \citenamefont {Cardoso}, \citenamefont {Carullo} \emph
  {et~al.}}]{BertiEtAl2026}%
  \BibitemOpen
  \bibfield  {author} {\bibinfo {author} {\bibfnamefont {E.}~\bibnamefont
  {Berti}}, \bibinfo {author} {\bibfnamefont {V.}~\bibnamefont {Cardoso}},
  \bibinfo {author} {\bibfnamefont {G.}~\bibnamefont {Carullo}}, \emph
  {et~al.},\ }\bibfield  {title} {\bibinfo {title} {Black hole spectroscopy:
  From theory to experiment},\ }\href
  {https://doi.org/10.1088/1361-6382/ae59e2} {\bibfield  {journal} {\bibinfo
  {journal} {Class. Quantum Grav.}\ }\textbf {\bibinfo {volume} {43}},\
  \bibinfo {pages} {123001} (\bibinfo {year} {2026})}\BibitemShut {NoStop}%
\bibitem [{\citenamefont {Ay\'on-Beato}\ and\ \citenamefont
  {Garc\'ia}(1998)}]{AyonBeatoGarcia1998}%
  \BibitemOpen
  \bibfield  {author} {\bibinfo {author} {\bibfnamefont {E.}~\bibnamefont
  {Ay\'on-Beato}}\ and\ \bibinfo {author} {\bibfnamefont {A.}~\bibnamefont
  {Garc\'ia}},\ }\bibfield  {title} {\bibinfo {title} {Regular black hole in
  general relativity coupled to nonlinear electrodynamics},\ }\href
  {https://doi.org/10.1103/PhysRevLett.80.5056} {\bibfield  {journal} {\bibinfo
   {journal} {Phys. Rev. Lett.}\ }\textbf {\bibinfo {volume} {80}},\ \bibinfo
  {pages} {5056} (\bibinfo {year} {1998})}\BibitemShut {NoStop}%
\bibitem [{\citenamefont {Hayward}(2006)}]{Hayward2006}%
  \BibitemOpen
  \bibfield  {author} {\bibinfo {author} {\bibfnamefont {S.~A.}\ \bibnamefont
  {Hayward}},\ }\bibfield  {title} {\bibinfo {title} {Formation and evaporation
  of nonsingular black holes},\ }\href
  {https://doi.org/10.1103/PhysRevLett.96.031103} {\bibfield  {journal}
  {\bibinfo  {journal} {Phys. Rev. Lett.}\ }\textbf {\bibinfo {volume} {96}},\
  \bibinfo {pages} {031103} (\bibinfo {year} {2006})}\BibitemShut {NoStop}%
\bibitem [{\citenamefont {Bronnikov}(2001)}]{Bronnikov2001}%
  \BibitemOpen
  \bibfield  {author} {\bibinfo {author} {\bibfnamefont {K.~A.}\ \bibnamefont
  {Bronnikov}},\ }\bibfield  {title} {\bibinfo {title} {Regular magnetic black
  holes and monopoles from nonlinear electrodynamics},\ }\href
  {https://doi.org/10.1103/PhysRevD.63.044005} {\bibfield  {journal} {\bibinfo
  {journal} {Phys. Rev. D}\ }\textbf {\bibinfo {volume} {63}},\ \bibinfo
  {pages} {044005} (\bibinfo {year} {2001})}\BibitemShut {NoStop}%
\bibitem [{\citenamefont {Dutta~Roy}\ and\ \citenamefont
  {Kar}(2022)}]{DuttaRoyKar2022}%
  \BibitemOpen
  \bibfield  {author} {\bibinfo {author} {\bibfnamefont {P.}~\bibnamefont
  {Dutta~Roy}}\ and\ \bibinfo {author} {\bibfnamefont {S.}~\bibnamefont
  {Kar}},\ }\bibfield  {title} {\bibinfo {title} {Generalized hayward
  spacetimes: Geometry, matter, and scalar quasinormal modes},\ }\href
  {https://doi.org/10.1103/PhysRevD.106.044028} {\bibfield  {journal} {\bibinfo
   {journal} {Phys. Rev. D}\ }\textbf {\bibinfo {volume} {106}},\ \bibinfo
  {pages} {044028} (\bibinfo {year} {2022})}\BibitemShut {NoStop}%
\bibitem [{\citenamefont {Kudryavcev}\ \emph {et~al.}(2026)\citenamefont
  {Kudryavcev}, \citenamefont {Ling},\ and\ \citenamefont
  {Vertogradov}}]{KudryavcevLingVertogradov2026}%
  \BibitemOpen
  \bibfield  {author} {\bibinfo {author} {\bibfnamefont {D.}~\bibnamefont
  {Kudryavcev}}, \bibinfo {author} {\bibfnamefont {Y.}~\bibnamefont {Ling}},\
  and\ \bibinfo {author} {\bibfnamefont {V.}~\bibnamefont {Vertogradov}},\
  }\bibfield  {title} {\bibinfo {title} {Exact dynamical regular black holes
  from generalized polytropic matter},\ }\href
  {https://doi.org/10.1140/epjc/s10052-026-15869-9} {\bibfield  {journal}
  {\bibinfo  {journal} {Eur. Phys. J. C}\ }\textbf {\bibinfo {volume} {86}},\
  \bibinfo {pages} {780} (\bibinfo {year} {2026})}\BibitemShut {NoStop}%
\bibitem [{\citenamefont {Ay\'on-Beato}\ and\ \citenamefont
  {Garc\'ia}(2000)}]{AyonBeatoGarcia2000}%
  \BibitemOpen
  \bibfield  {author} {\bibinfo {author} {\bibfnamefont {E.}~\bibnamefont
  {Ay\'on-Beato}}\ and\ \bibinfo {author} {\bibfnamefont {A.}~\bibnamefont
  {Garc\'ia}},\ }\bibfield  {title} {\bibinfo {title} {The bardeen model as a
  nonlinear magnetic monopole},\ }\href
  {https://doi.org/10.1016/S0370-2693(00)01125-4} {\bibfield  {journal}
  {\bibinfo  {journal} {Phys. Lett. B}\ }\textbf {\bibinfo {volume} {493}},\
  \bibinfo {pages} {149} (\bibinfo {year} {2000})}\BibitemShut {NoStop}%
\bibitem [{\citenamefont {Moreno}\ and\ \citenamefont
  {Sarbach}(2003)}]{MorenoSarbach2003}%
  \BibitemOpen
  \bibfield  {author} {\bibinfo {author} {\bibfnamefont {C.}~\bibnamefont
  {Moreno}}\ and\ \bibinfo {author} {\bibfnamefont {O.}~\bibnamefont
  {Sarbach}},\ }\bibfield  {title} {\bibinfo {title} {Stability properties of
  black holes in self-gravitating nonlinear electrodynamics},\ }\href
  {https://doi.org/10.1103/PhysRevD.67.024028} {\bibfield  {journal} {\bibinfo
  {journal} {Phys. Rev. D}\ }\textbf {\bibinfo {volume} {67}},\ \bibinfo
  {pages} {024028} (\bibinfo {year} {2003})}\BibitemShut {NoStop}%
\bibitem [{\citenamefont {Novello}\ \emph {et~al.}(2000)\citenamefont
  {Novello}, \citenamefont {De~Lorenci}, \citenamefont {Salim},\ and\
  \citenamefont {Klippert}}]{NovelloEtAl2000}%
  \BibitemOpen
  \bibfield  {author} {\bibinfo {author} {\bibfnamefont {M.}~\bibnamefont
  {Novello}}, \bibinfo {author} {\bibfnamefont {V.~A.}\ \bibnamefont
  {De~Lorenci}}, \bibinfo {author} {\bibfnamefont {J.~M.}\ \bibnamefont
  {Salim}},\ and\ \bibinfo {author} {\bibfnamefont {R.}~\bibnamefont
  {Klippert}},\ }\bibfield  {title} {\bibinfo {title} {Geometrical aspects of
  light propagation in nonlinear electrodynamics},\ }\href
  {https://doi.org/10.1103/PhysRevD.61.045001} {\bibfield  {journal} {\bibinfo
  {journal} {Phys. Rev. D}\ }\textbf {\bibinfo {volume} {61}},\ \bibinfo
  {pages} {045001} (\bibinfo {year} {2000})}\BibitemShut {NoStop}%
\bibitem [{\citenamefont {Schellstede}\ \emph {et~al.}(2016)\citenamefont
  {Schellstede}, \citenamefont {Perlick},\ and\ \citenamefont
  {L\"ammerzahl}}]{SchellstedeEtAl2016}%
  \BibitemOpen
  \bibfield  {author} {\bibinfo {author} {\bibfnamefont {G.~O.}\ \bibnamefont
  {Schellstede}}, \bibinfo {author} {\bibfnamefont {V.}~\bibnamefont
  {Perlick}},\ and\ \bibinfo {author} {\bibfnamefont {C.}~\bibnamefont
  {L\"ammerzahl}},\ }\bibfield  {title} {\bibinfo {title} {On causality in
  nonlinear vacuum electrodynamics of the pleba\'nski class},\ }\href
  {https://doi.org/10.1002/andp.201600124} {\bibfield  {journal} {\bibinfo
  {journal} {Ann. Phys. (Berlin)}\ }\textbf {\bibinfo {volume} {528}},\
  \bibinfo {pages} {738} (\bibinfo {year} {2016})}\BibitemShut {NoStop}%
\bibitem [{\citenamefont {Toshmatov}\ \emph
  {et~al.}(2018{\natexlab{a}})\citenamefont {Toshmatov}, \citenamefont
  {Stuchl\'ik}, \citenamefont {Schee},\ and\ \citenamefont
  {Ahmedov}}]{ToshmatovEtAl2018}%
  \BibitemOpen
  \bibfield  {author} {\bibinfo {author} {\bibfnamefont {B.}~\bibnamefont
  {Toshmatov}}, \bibinfo {author} {\bibfnamefont {Z.}~\bibnamefont
  {Stuchl\'ik}}, \bibinfo {author} {\bibfnamefont {J.}~\bibnamefont {Schee}},\
  and\ \bibinfo {author} {\bibfnamefont {B.}~\bibnamefont {Ahmedov}},\
  }\bibfield  {title} {\bibinfo {title} {Electromagnetic perturbations of black
  holes in general relativity coupled to nonlinear electrodynamics},\ }\href
  {https://doi.org/10.1103/PhysRevD.97.084058} {\bibfield  {journal} {\bibinfo
  {journal} {Phys. Rev. D}\ }\textbf {\bibinfo {volume} {97}},\ \bibinfo
  {pages} {084058} (\bibinfo {year} {2018}{\natexlab{a}})}\BibitemShut
  {NoStop}%
\bibitem [{\citenamefont {Toshmatov}\ \emph
  {et~al.}(2018{\natexlab{b}})\citenamefont {Toshmatov}, \citenamefont
  {Stuchl\'ik},\ and\ \citenamefont {Ahmedov}}]{ToshmatovEtAl2018Polar}%
  \BibitemOpen
  \bibfield  {author} {\bibinfo {author} {\bibfnamefont {B.}~\bibnamefont
  {Toshmatov}}, \bibinfo {author} {\bibfnamefont {Z.}~\bibnamefont
  {Stuchl\'ik}},\ and\ \bibinfo {author} {\bibfnamefont {B.}~\bibnamefont
  {Ahmedov}},\ }\bibfield  {title} {\bibinfo {title} {Electromagnetic
  perturbations of black holes in general relativity coupled to nonlinear
  electrodynamics: Polar perturbations},\ }\href
  {https://doi.org/10.1103/PhysRevD.98.085021} {\bibfield  {journal} {\bibinfo
  {journal} {Phys. Rev. D}\ }\textbf {\bibinfo {volume} {98}},\ \bibinfo
  {pages} {085021} (\bibinfo {year} {2018}{\natexlab{b}})}\BibitemShut
  {NoStop}%
\bibitem [{\citenamefont {Tomizawa}\ and\ \citenamefont
  {Suzuki}(2023)}]{TomizawaSuzuki2023}%
  \BibitemOpen
  \bibfield  {author} {\bibinfo {author} {\bibfnamefont {S.}~\bibnamefont
  {Tomizawa}}\ and\ \bibinfo {author} {\bibfnamefont {R.}~\bibnamefont
  {Suzuki}},\ }\bibfield  {title} {\bibinfo {title} {Causality of photon
  propagation under dominant energy condition in nonlinear electrodynamics},\
  }\href {https://doi.org/10.1103/PhysRevD.108.124072} {\bibfield  {journal}
  {\bibinfo  {journal} {Phys. Rev. D}\ }\textbf {\bibinfo {volume} {108}},\
  \bibinfo {pages} {124072} (\bibinfo {year} {2023})}\BibitemShut {NoStop}%
\bibitem [{\citenamefont {Daghigh}\ and\ \citenamefont
  {Green}(2022)}]{DaghighGreen2022}%
  \BibitemOpen
  \bibfield  {author} {\bibinfo {author} {\bibfnamefont {R.~G.}\ \bibnamefont
  {Daghigh}}\ and\ \bibinfo {author} {\bibfnamefont {M.~D.}\ \bibnamefont
  {Green}},\ }\bibfield  {title} {\bibinfo {title} {Gravitational and
  electromagnetic radiation from an electrically charged black hole in general
  nonlinear electrodynamics},\ }\href
  {https://doi.org/10.1103/PhysRevD.105.024055} {\bibfield  {journal} {\bibinfo
   {journal} {Phys. Rev. D}\ }\textbf {\bibinfo {volume} {105}},\ \bibinfo
  {pages} {024055} (\bibinfo {year} {2022})}\BibitemShut {NoStop}%
\bibitem [{\citenamefont {Bret\'on}(2005)}]{Breton2005}%
  \BibitemOpen
  \bibfield  {author} {\bibinfo {author} {\bibfnamefont {N.}~\bibnamefont
  {Bret\'on}},\ }\bibfield  {title} {\bibinfo {title} {Stability of nonlinear
  magnetic black holes},\ }\href {https://doi.org/10.1103/PhysRevD.72.044015}
  {\bibfield  {journal} {\bibinfo  {journal} {Phys. Rev. D}\ }\textbf {\bibinfo
  {volume} {72}},\ \bibinfo {pages} {044015} (\bibinfo {year}
  {2005})}\BibitemShut {NoStop}%
\bibitem [{\citenamefont {Nomura}\ \emph {et~al.}(2020)\citenamefont {Nomura},
  \citenamefont {Yoshida},\ and\ \citenamefont {Soda}}]{NomuraEtAl2020}%
  \BibitemOpen
  \bibfield  {author} {\bibinfo {author} {\bibfnamefont {K.}~\bibnamefont
  {Nomura}}, \bibinfo {author} {\bibfnamefont {D.}~\bibnamefont {Yoshida}},\
  and\ \bibinfo {author} {\bibfnamefont {J.}~\bibnamefont {Soda}},\ }\bibfield
  {title} {\bibinfo {title} {Stability of magnetic black holes in general
  nonlinear electrodynamics},\ }\href
  {https://doi.org/10.1103/PhysRevD.101.124026} {\bibfield  {journal} {\bibinfo
   {journal} {Phys. Rev. D}\ }\textbf {\bibinfo {volume} {101}},\ \bibinfo
  {pages} {124026} (\bibinfo {year} {2020})}\BibitemShut {NoStop}%
\bibitem [{\citenamefont {Chaverra}\ \emph {et~al.}(2016)\citenamefont
  {Chaverra}, \citenamefont {Degollado}, \citenamefont {Moreno},\ and\
  \citenamefont {Sarbach}}]{ChaverraEtAl2016}%
  \BibitemOpen
  \bibfield  {author} {\bibinfo {author} {\bibfnamefont {E.}~\bibnamefont
  {Chaverra}}, \bibinfo {author} {\bibfnamefont {J.~C.}\ \bibnamefont
  {Degollado}}, \bibinfo {author} {\bibfnamefont {C.}~\bibnamefont {Moreno}},\
  and\ \bibinfo {author} {\bibfnamefont {O.}~\bibnamefont {Sarbach}},\
  }\bibfield  {title} {\bibinfo {title} {Black holes in nonlinear
  electrodynamics: Quasinormal spectra and parity splitting},\ }\href
  {https://doi.org/10.1103/PhysRevD.93.123013} {\bibfield  {journal} {\bibinfo
  {journal} {Phys. Rev. D}\ }\textbf {\bibinfo {volume} {93}},\ \bibinfo
  {pages} {123013} (\bibinfo {year} {2016})}\BibitemShut {NoStop}%
\bibitem [{\citenamefont {Nomura}\ and\ \citenamefont
  {Yoshida}(2022)}]{NomuraYoshida2022}%
  \BibitemOpen
  \bibfield  {author} {\bibinfo {author} {\bibfnamefont {K.}~\bibnamefont
  {Nomura}}\ and\ \bibinfo {author} {\bibfnamefont {D.}~\bibnamefont
  {Yoshida}},\ }\bibfield  {title} {\bibinfo {title} {Quasinormal modes of
  charged black holes with corrections from nonlinear electrodynamics},\ }\href
  {https://doi.org/10.1103/PhysRevD.105.044006} {\bibfield  {journal} {\bibinfo
   {journal} {Phys. Rev. D}\ }\textbf {\bibinfo {volume} {105}},\ \bibinfo
  {pages} {044006} (\bibinfo {year} {2022})}\BibitemShut {NoStop}%
\bibitem [{\citenamefont {Meng}\ and\ \citenamefont
  {Zhang}(2023)}]{MengZhang2023}%
  \BibitemOpen
  \bibfield  {author} {\bibinfo {author} {\bibfnamefont {K.}~\bibnamefont
  {Meng}}\ and\ \bibinfo {author} {\bibfnamefont {S.-J.}\ \bibnamefont
  {Zhang}},\ }\bibfield  {title} {\bibinfo {title} {Gravito-electromagnetic
  perturbations and qnms of regular black holes},\ }\href
  {https://doi.org/10.1088/1361-6382/acf3c6} {\bibfield  {journal} {\bibinfo
  {journal} {Class. Quantum Grav.}\ }\textbf {\bibinfo {volume} {40}},\
  \bibinfo {pages} {195024} (\bibinfo {year} {2023})}\BibitemShut {NoStop}%
\bibitem [{\citenamefont {Malik}(2025)}]{Malik2025}%
  \BibitemOpen
  \bibfield  {author} {\bibinfo {author} {\bibfnamefont {Z.}~\bibnamefont
  {Malik}},\ }\bibfield  {title} {\bibinfo {title} {Gravitational perturbations
  of the hayward spacetime and testing the correspondence between quasinormal
  modes and grey-body factors},\ }\href
  {https://doi.org/10.1007/s10773-025-06198-w} {\bibfield  {journal} {\bibinfo
  {journal} {Int. J. Theor. Phys.}\ }\textbf {\bibinfo {volume} {64}},\
  \bibinfo {pages} {314} (\bibinfo {year} {2025})}\BibitemShut {NoStop}%
\bibitem [{\citenamefont {Wu}\ \emph {et~al.}(2025)\citenamefont {Wu},
  \citenamefont {Cai},\ and\ \citenamefont {Xie}}]{WuCaiXie2025}%
  \BibitemOpen
  \bibfield  {author} {\bibinfo {author} {\bibfnamefont {L.-B.}\ \bibnamefont
  {Wu}}, \bibinfo {author} {\bibfnamefont {R.-G.}\ \bibnamefont {Cai}},\ and\
  \bibinfo {author} {\bibfnamefont {L.}~\bibnamefont {Xie}},\ }\bibfield
  {title} {\bibinfo {title} {Stability of the greybody factor of hayward black
  holes},\ }\href {https://doi.org/10.1103/PhysRevD.111.044066} {\bibfield
  {journal} {\bibinfo  {journal} {Phys. Rev. D}\ }\textbf {\bibinfo {volume}
  {111}},\ \bibinfo {pages} {044066} (\bibinfo {year} {2025})}\BibitemShut
  {NoStop}%
\bibitem [{\citenamefont {Bolokhov}\ and\ \citenamefont
  {Skvortsova}(2026)}]{BolokhovSkvortsova2026}%
  \BibitemOpen
  \bibfield  {author} {\bibinfo {author} {\bibfnamefont {S.~V.}\ \bibnamefont
  {Bolokhov}}\ and\ \bibinfo {author} {\bibfnamefont {M.}~\bibnamefont
  {Skvortsova}},\ }\bibfield  {title} {\bibinfo {title} {Gravitational
  quasinormal modes of the hayward spacetime},\ }\href
  {https://doi.org/10.1140/epjc/s10052-026-15624-0} {\bibfield  {journal}
  {\bibinfo  {journal} {Eur. Phys. J. C}\ }\textbf {\bibinfo {volume} {86}},\
  \bibinfo {pages} {374} (\bibinfo {year} {2026})}\BibitemShut {NoStop}%
\bibitem [{\citenamefont {Liang}\ \emph {et~al.}(2026)\citenamefont {Liang},
  \citenamefont {Liu},\ and\ \citenamefont {Long}}]{LiangEtAl2026}%
  \BibitemOpen
  \bibfield  {author} {\bibinfo {author} {\bibfnamefont {J.}~\bibnamefont
  {Liang}}, \bibinfo {author} {\bibfnamefont {D.}~\bibnamefont {Liu}},\ and\
  \bibinfo {author} {\bibfnamefont {Z.-W.}\ \bibnamefont {Long}},\ }\bibfield
  {title} {\bibinfo {title} {Quasinormal modes and greybody factors of black
  holes corrected by nonlinear electrodynamics},\ }\href
  {https://doi.org/10.1140/epjc/s10052-025-15245-z} {\bibfield  {journal}
  {\bibinfo  {journal} {Eur. Phys. J. C}\ }\textbf {\bibinfo {volume} {86}},\
  \bibinfo {pages} {17} (\bibinfo {year} {2026})}\BibitemShut {NoStop}%
\bibitem [{\citenamefont {Ould El~Hadj}\ and\ \citenamefont
  {Dolan}(2022)}]{OuldElHadjDolan2022}%
  \BibitemOpen
  \bibfield  {author} {\bibinfo {author} {\bibfnamefont {M.}~\bibnamefont {Ould
  El~Hadj}}\ and\ \bibinfo {author} {\bibfnamefont {S.~R.}\ \bibnamefont
  {Dolan}},\ }\bibfield  {title} {\bibinfo {title} {Conversion of
  electromagnetic and gravitational waves by a charged black hole},\ }\href
  {https://doi.org/10.1103/PhysRevD.106.044002} {\bibfield  {journal} {\bibinfo
   {journal} {Phys. Rev. D}\ }\textbf {\bibinfo {volume} {106}},\ \bibinfo
  {pages} {044002} (\bibinfo {year} {2022})}\BibitemShut {NoStop}%
\bibitem [{\citenamefont {De~Felice}\ and\ \citenamefont
  {Tsujikawa}(2025)}]{DeFeliceTsujikawa2025}%
  \BibitemOpen
  \bibfield  {author} {\bibinfo {author} {\bibfnamefont {A.}~\bibnamefont
  {De~Felice}}\ and\ \bibinfo {author} {\bibfnamefont {S.}~\bibnamefont
  {Tsujikawa}},\ }\bibfield  {title} {\bibinfo {title} {Instability of
  nonsingular black holes in nonlinear electrodynamics},\ }\href
  {https://doi.org/10.1103/PhysRevLett.134.081401} {\bibfield  {journal}
  {\bibinfo  {journal} {Phys. Rev. Lett.}\ }\textbf {\bibinfo {volume} {134}},\
  \bibinfo {pages} {081401} (\bibinfo {year} {2025})}\BibitemShut {NoStop}%
\bibitem [{\citenamefont {Huang}\ and\ \citenamefont
  {Rao}(2025)}]{HuangRao2025}%
  \BibitemOpen
  \bibfield  {author} {\bibinfo {author} {\bibfnamefont {H.}~\bibnamefont
  {Huang}}\ and\ \bibinfo {author} {\bibfnamefont {X.-P.}\ \bibnamefont
  {Rao}},\ }\bibfield  {title} {\bibinfo {title} {Regular black holes and their
  singular families},\ }\href {https://doi.org/10.1103/PhysRevD.111.104040}
  {\bibfield  {journal} {\bibinfo  {journal} {Phys. Rev. D}\ }\textbf {\bibinfo
  {volume} {111}},\ \bibinfo {pages} {104040} (\bibinfo {year}
  {2025})}\BibitemShut {NoStop}%
\bibitem [{\citenamefont {Bokuli\'c}\ \emph {et~al.}(2026)\citenamefont
  {Bokuli\'c}, \citenamefont {Juri\'c},\ and\ \citenamefont
  {Smoli\'c}}]{BokulicJuricSmolic2026}%
  \BibitemOpen
  \bibfield  {author} {\bibinfo {author} {\bibfnamefont {A.}~\bibnamefont
  {Bokuli\'c}}, \bibinfo {author} {\bibfnamefont {T.}~\bibnamefont
  {Juri\'c}},\ and\ \bibinfo {author} {\bibfnamefont {I.}~\bibnamefont
  {Smoli\'c}},\ }\bibfield  {title} {\bibinfo {title} {Conundrum of regular
  black holes with nonlinear electromagnetic fields},\ }\href
  {https://doi.org/10.1103/z7gd-96ms} {\bibfield  {journal} {\bibinfo
  {journal} {Phys. Rev. D}\ }\textbf {\bibinfo {volume} {113}},\ \bibinfo
  {pages} {024044} (\bibinfo {year} {2026})}\BibitemShut {NoStop}%
\bibitem [{\citenamefont {Moncrief}(1974{\natexlab{a}})}]{Moncrief1974a}%
  \BibitemOpen
  \bibfield  {author} {\bibinfo {author} {\bibfnamefont {V.}~\bibnamefont
  {Moncrief}},\ }\bibfield  {title} {\bibinfo {title} {Odd-parity stability of
  a reissner--nordstr\"om black hole},\ }\href
  {https://doi.org/10.1103/PhysRevD.9.2707} {\bibfield  {journal} {\bibinfo
  {journal} {Phys. Rev. D}\ }\textbf {\bibinfo {volume} {9}},\ \bibinfo {pages}
  {2707} (\bibinfo {year} {1974}{\natexlab{a}})}\BibitemShut {NoStop}%
\bibitem [{\citenamefont {Moncrief}(1974{\natexlab{b}})}]{Moncrief1974b}%
  \BibitemOpen
  \bibfield  {author} {\bibinfo {author} {\bibfnamefont {V.}~\bibnamefont
  {Moncrief}},\ }\bibfield  {title} {\bibinfo {title} {Stability of
  reissner--nordstr\"om black holes},\ }\href
  {https://doi.org/10.1103/PhysRevD.10.1057} {\bibfield  {journal} {\bibinfo
  {journal} {Phys. Rev. D}\ }\textbf {\bibinfo {volume} {10}},\ \bibinfo
  {pages} {1057} (\bibinfo {year} {1974}{\natexlab{b}})}\BibitemShut {NoStop}%
\bibitem [{\citenamefont {Leaver}(1985)}]{Leaver1985}%
  \BibitemOpen
  \bibfield  {author} {\bibinfo {author} {\bibfnamefont {E.~W.}\ \bibnamefont
  {Leaver}},\ }\bibfield  {title} {\bibinfo {title} {An analytic representation
  for the quasinormal modes of kerr black holes},\ }\href
  {https://doi.org/10.1098/rspa.1985.0119} {\bibfield  {journal} {\bibinfo
  {journal} {Proc. R. Soc. A}\ }\textbf {\bibinfo {volume} {402}},\ \bibinfo
  {pages} {285} (\bibinfo {year} {1985})}\BibitemShut {NoStop}%
\bibitem [{\citenamefont {Jansen}(2017)}]{Jansen2017}%
  \BibitemOpen
  \bibfield  {author} {\bibinfo {author} {\bibfnamefont {A.}~\bibnamefont
  {Jansen}},\ }\bibfield  {title} {\bibinfo {title} {Overdamped modes in
  schwarzschild de sitter and a mathematica package for the numerical
  computation of quasinormal modes},\ }\href
  {https://doi.org/10.1140/epjp/i2017-11825-9} {\bibfield  {journal} {\bibinfo
  {journal} {Eur. Phys. J. Plus}\ }\textbf {\bibinfo {volume} {132}},\ \bibinfo
  {pages} {546} (\bibinfo {year} {2017})}\BibitemShut {NoStop}%
\bibitem [{\citenamefont {Gundlach}\ \emph {et~al.}(1994)\citenamefont
  {Gundlach}, \citenamefont {Price},\ and\ \citenamefont
  {Pullin}}]{GundlachPricePullin1994}%
  \BibitemOpen
  \bibfield  {author} {\bibinfo {author} {\bibfnamefont {C.}~\bibnamefont
  {Gundlach}}, \bibinfo {author} {\bibfnamefont {R.~H.}\ \bibnamefont
  {Price}},\ and\ \bibinfo {author} {\bibfnamefont {J.}~\bibnamefont
  {Pullin}},\ }\bibfield  {title} {\bibinfo {title} {Late-time behavior of
  stellar collapse and explosions. i. linearized perturbations},\ }\href
  {https://doi.org/10.1103/PhysRevD.49.883} {\bibfield  {journal} {\bibinfo
  {journal} {Phys. Rev. D}\ }\textbf {\bibinfo {volume} {49}},\ \bibinfo
  {pages} {883} (\bibinfo {year} {1994})}\BibitemShut {NoStop}%
\bibitem [{\citenamefont {Cardoso}\ \emph {et~al.}(2009)\citenamefont
  {Cardoso}, \citenamefont {Miranda}, \citenamefont {Berti}, \citenamefont
  {Witek},\ and\ \citenamefont {Zanchin}}]{CardosoEtAl2009}%
  \BibitemOpen
  \bibfield  {author} {\bibinfo {author} {\bibfnamefont {V.}~\bibnamefont
  {Cardoso}}, \bibinfo {author} {\bibfnamefont {A.~S.}\ \bibnamefont
  {Miranda}}, \bibinfo {author} {\bibfnamefont {E.}~\bibnamefont {Berti}},
  \bibinfo {author} {\bibfnamefont {H.}~\bibnamefont {Witek}},\ and\ \bibinfo
  {author} {\bibfnamefont {V.~T.}\ \bibnamefont {Zanchin}},\ }\bibfield
  {title} {\bibinfo {title} {Geodesic stability, lyapunov exponents, and
  quasinormal modes},\ }\href {https://doi.org/10.1103/PhysRevD.79.064016}
  {\bibfield  {journal} {\bibinfo  {journal} {Phys. Rev. D}\ }\textbf {\bibinfo
  {volume} {79}},\ \bibinfo {pages} {064016} (\bibinfo {year}
  {2009})}\BibitemShut {NoStop}%
\bibitem [{\citenamefont {Aretakis}(2011)}]{Aretakis2011}%
  \BibitemOpen
  \bibfield  {author} {\bibinfo {author} {\bibfnamefont {S.}~\bibnamefont
  {Aretakis}},\ }\bibfield  {title} {\bibinfo {title} {Stability and
  instability of extreme reissner--nordstr\"om black hole spacetimes for linear
  scalar perturbations i},\ }\href {https://doi.org/10.1007/s00220-011-1254-5}
  {\bibfield  {journal} {\bibinfo  {journal} {Commun. Math. Phys.}\ }\textbf
  {\bibinfo {volume} {307}},\ \bibinfo {pages} {17} (\bibinfo {year}
  {2011})}\BibitemShut {NoStop}%
\bibitem [{\citenamefont {Lucietti}\ and\ \citenamefont
  {Reall}(2012)}]{LuciettiReall2012}%
  \BibitemOpen
  \bibfield  {author} {\bibinfo {author} {\bibfnamefont {J.}~\bibnamefont
  {Lucietti}}\ and\ \bibinfo {author} {\bibfnamefont {H.~S.}\ \bibnamefont
  {Reall}},\ }\bibfield  {title} {\bibinfo {title} {Gravitational instability
  of an extreme kerr black hole},\ }\href
  {https://doi.org/10.1103/PhysRevD.86.104030} {\bibfield  {journal} {\bibinfo
  {journal} {Phys. Rev. D}\ }\textbf {\bibinfo {volume} {86}},\ \bibinfo
  {pages} {104030} (\bibinfo {year} {2012})}\BibitemShut {NoStop}%
\bibitem [{\citenamefont {Berti}\ \emph {et~al.}(2006)\citenamefont {Berti},
  \citenamefont {Cardoso},\ and\ \citenamefont {Will}}]{BertiCardosoWill2006}%
  \BibitemOpen
  \bibfield  {author} {\bibinfo {author} {\bibfnamefont {E.}~\bibnamefont
  {Berti}}, \bibinfo {author} {\bibfnamefont {V.}~\bibnamefont {Cardoso}},\
  and\ \bibinfo {author} {\bibfnamefont {C.~M.}\ \bibnamefont {Will}},\
  }\bibfield  {title} {\bibinfo {title} {Gravitational wave spectroscopy of
  massive black holes with the space interferometer lisa},\ }\href
  {https://doi.org/10.1103/PhysRevD.73.064030} {\bibfield  {journal} {\bibinfo
  {journal} {Phys. Rev. D}\ }\textbf {\bibinfo {volume} {73}},\ \bibinfo
  {pages} {064030} (\bibinfo {year} {2006})}\BibitemShut {NoStop}%
\bibitem [{\citenamefont {Silva}\ \emph {et~al.}(2024)\citenamefont {Silva},
  \citenamefont {Tambalo}, \citenamefont {Glampedakis}, \citenamefont {Yagi},\
  and\ \citenamefont {Steinhoff}}]{SilvaEtAl2024}%
  \BibitemOpen
  \bibfield  {author} {\bibinfo {author} {\bibfnamefont {H.~O.}\ \bibnamefont
  {Silva}}, \bibinfo {author} {\bibfnamefont {G.}~\bibnamefont {Tambalo}},
  \bibinfo {author} {\bibfnamefont {K.}~\bibnamefont {Glampedakis}}, \bibinfo
  {author} {\bibfnamefont {K.}~\bibnamefont {Yagi}},\ and\ \bibinfo {author}
  {\bibfnamefont {J.}~\bibnamefont {Steinhoff}},\ }\bibfield  {title} {\bibinfo
  {title} {Quasinormal modes and their excitation beyond general relativity},\
  }\href {https://doi.org/10.1103/PhysRevD.110.024042} {\bibfield  {journal}
  {\bibinfo  {journal} {Phys. Rev. D}\ }\textbf {\bibinfo {volume} {110}},\
  \bibinfo {pages} {024042} (\bibinfo {year} {2024})}\BibitemShut {NoStop}%
\end{thebibliography}
\end{document}